# Quantum Fourier states and gates: Teleportation via rough entanglement

Mario Mastriani

*Knight Foundation School of Computing & Information Sciences,*
*Florida International University, 11200 S.W. 8th Street, Miami, FL 33199, USA*

Quantum Fourier gates (QFG) constitute a family of quantum gates that result from an exact combination of the quantum Fourier transform (QFT) and the SWAP gate. As a direct consequence of this, the Feynman gate is a particular case of that family, just as the Bell states are particular cases of the states that are also derived from the aforementioned family. Besides, this new tool will allow us to demonstrate that teleportation is not something that happens exclusively thanks to some level of entanglement, but that it is also possible with an incomplete form of entanglement known as rough entanglement. Finally, other applications necessary in the quantum Internet are incorporated.

ORCID Id: 0000-0002-5627-3935
mmastria@fiu.edu

***Introduction***.- During the last almost two and a half centuries, there have been innumerable interventions of Fourier's works (Jean-Baptiste Joseph Fourier, Auxerre, France, March 21, 1768, Paris, May 16, 1830) in fields of Science as diverse as Mathematics, Physics, Chemistry, Electronics, Bioengineering, and more recently Quantum Information Processing (QIP)[1]. In the latter case, one application stands out above all others, known as phase estimation[1], which is of vital importance in order-finding[1] and factoring algorithms[1]. The most famous quantum algorithm in the QIP arsenal is without a doubt Shor's algorithm[2], which consists of a hybrid configuration (classical-quantum) that is useful to factor an extremely large number into two other prime numbers in polynomial time.

Knowledge about the link between all QIP gates with QFT is relatively recent[3-6], specifically, we refer to QFT as an underlying generator of the aforementioned gates. In this sense, the alternative expressions of the Feynman gates (Control-X, CNOT, or simply CX), Toffoli[3], and Hadamard from the QFT stand out, being the Hadamard gate the same QFT for the case of a single qubit[3-6].

Moreover, in these works[3-6] the spectral nature of the entanglement was established, where both the Bell states[1] and the Greenberger-Horne-Zeilinger (GHZ) type configurations[1] arise from appropriate combinations of the QFT. In this way, the dual nature of entanglement was revealed, where the spectral aspect was added to its well-known temporal facet. Taking into account both sides of the entanglement will allow the development of new and better Quantum Communications protocols[7], in particular, more efficient implementations of the quantum teleportation protocol[8] with a strong projection on the future quantum Internet[9-14], given that when we try to implement quantum key distribution protocols on the ground[15] through fiber-optic lines, we must use quantum repeaters at regular distances[16]. The problem with these repeaters is that the key to be distributed is exposed when passing through them, so an alternative to this serious security problem is quantum teleportation[8]. Hence the importance of testing better implementations of this protocol[8] thanks to a better understanding of the inner springs of entanglement.

Finally, this work comes to fill interstitial and complementary spaces to those already mentioned in regards to our knowledge about entanglement and quantum teleportation, given that as will be demonstrated in this work, the essential element for teleportation is not entanglement but states derived from the application of QFT, of which entanglement is only a particular case.

***Quantum Fourier gates***.- As it was previously mentioned, quantum Fourier gates (QFG) constitute a family of quantum gates that result from an exact combination of the quantum Fourier transform (QFT)[1] and the SWAP gate[1]. In their generic form, these gates will be represented as

$$F_p^d = SWAP_{2^p \times 2^p} \; QFT_{2^p \times 2^p}^d \; SWAP_{2^p \times 2^p}, \tag{1}$$

where $F$ is the corresponding QFG, the subscript $p$ indicates the number of qubits involved by the gate (although the subscript $2^p \times 2^p$ represents the dimension of the QFT and SWAP matrices), while the superscript $d$ represents the degree of the gate, which is equivalent to the number of QFT blocks that the mentioned gate involves. A QFG can only have four possible degrees (0, 1, 2, and 3) regardless of the "number of QFT blocks" it includes. Specifically, "number of QFT blocks" $\equiv d$ (*mod* 4), i.e., the integers "number of QFT blocks" and *d* are said to be congruent modulo 4, if there is an integer *k* such that "number of QFT blocks" - $d = k \times 4$. Congruence modulo 4 is a congruence relation, where the parentheses mean that (*mod* 4) applies to the entire equation, not just to the right-hand side (here *d*). As it will be seen below, this is because the collection of four cascaded QFT blocks is equivalent to the identity matrix[1]. Without losing generality, this property of the QFT will be proved for the case of two qubits, where it is known that[3-6]:

$$QFT_{2^2 \times 2^2} \; QFT_{2^2 \times 2^2} = CNOT_{flipped} = \begin{bmatrix} 1 & 0 & 0 & 0 \\ 0 & 0 & 0 & 1 \\ 0 & 0 & 1 & 0 \\ 0 & 1 & 0 & 0 \end{bmatrix} = SWAP_{2^2 \times 2^2} \; CNOT \; SWAP_{2^2 \times 2^2}, \qquad (2)$$

with:

$$QFT_{2^2 \times 2^2} = \frac{1}{2} \begin{bmatrix} 1 & 1 & 1 & 1 \\ 1 & i & -1 & -i \\ 1 & -1 & 1 & -1 \\ 1 & -i & -1 & i \end{bmatrix}, \text{ with } i = \sqrt{-1}, \qquad (3a)$$

$$SWAP_{2^2 \times 2^2} = \begin{bmatrix} 1 & 0 & 0 & 0 \\ 0 & 0 & 1 & 0 \\ 0 & 1 & 0 & 0 \\ 0 & 0 & 0 & 1 \end{bmatrix}, \text{ and} \qquad (3b)$$

$$CNOT = \begin{bmatrix} 1 & 0 & 0 & 0 \\ 0 & 1 & 0 & 0 \\ 0 & 0 & 0 & 1 \\ 0 & 0 & 1 & 0 \end{bmatrix}. \qquad (3c)$$

Then, being

$$SWAP_{2^2 \times 2^2} \; SWAP_{2^2 \times 2^2} = CNOT \; CNOT = I_{2^2 \times 2^2} = \begin{bmatrix} 1 & 0 & 0 & 0 \\ 0 & 1 & 0 & 0 \\ 0 & 0 & 1 & 0 \\ 0 & 0 & 0 & 1 \end{bmatrix}, \qquad (4)$$

since all quantum gates must be reversible[1] and taking into account Eq.(2), it turns out that,

$$\begin{aligned} QFT_{2^2 \times 2^2} \; QFT_{2^2 \times 2^2} \; QFT_{2^2 \times 2^2} \; QFT_{2^2 \times 2^2} &= CNOT_{flipped} \; CNOT_{flipped} \\ &= \left( SWAP_{2^2 \times 2^2} \; CNOT \; SWAP_{2^2 \times 2^2} \right)\left( SWAP_{2^2 \times 2^2} \; CNOT \; SWAP_{2^2 \times 2^2} \right) \\ &= I_{2^2 \times 2^2}. \end{aligned} \qquad (5)$$

Equation (5) indirectly proves the validity of the postulate "number of QFT blocks" $\equiv d$ (*mod* 4).

From now on, it is necessary to define a generalized SWAP gate for *p* qubits, called *swapping between equidistant qubits* (SBEQ) gate. Based on Fig. 1, *p* qubits will be numbered in ascending order, that is, from top to bottom of such figure, or what is similar, from 0 to (*p*-1). Therefore, ∀ *p* (even or odd) it turns out that the SBEQ gate performs swap operations between qubits equidistant from the center whose index is (*p*-1)/2, and in the correct order from the ends towards the mentioned center:

$$q[k] \leftrightarrow q[p-1-k], \forall k \in \{0, \lfloor (p-2)/2 \rfloor\}, \qquad (6)$$

where the operator $\lfloor \bullet \rfloor$ returns the smallest integer greater than or equal to the specified numeric expression "•", meanwhile, Fig. 1 shows a generic representation of the SBEQ gate for *p* qubits.

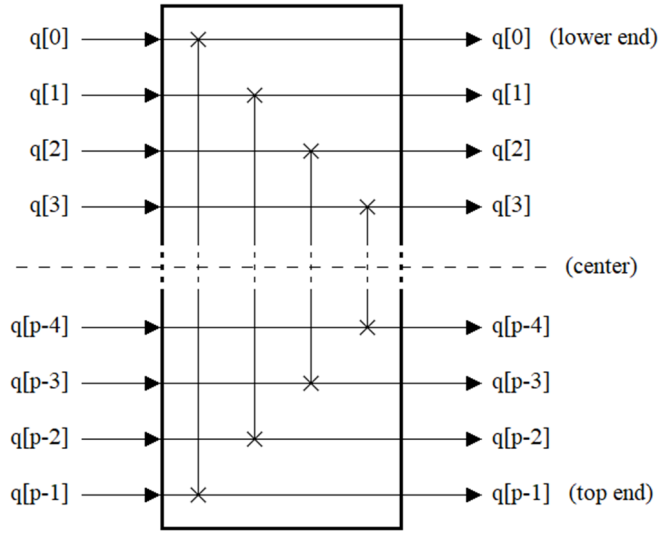

FIG. 1. Swapping-between-equidistant-qubits (SBEQ) gate for *p* qubits are exchanged. Individual exchanges occur between qubits opposite by the center, in the correct order from the ends towards the center, both for odd and even *p*.

The domain of *k* also represents the number of individual swaps that are applied within the SBEQ gate. For an odd *p*, the center qubit (i.e., *q*[(*p*-1)/2]) remains unchanged. A particular case of odd *p* occurs for *p* = 1, where the SBEQ gate implies only one SWAP gate, which leaves the only qubit intact since it is the identity matrix[1],

$$SWAP_{2^1 \times 2^1} = I_{2^1 \times 2^1} = \begin{bmatrix} 1 & 0 \\ 0 & 1 \end{bmatrix}. \qquad (7)$$

Equation (7) must be taken for what it is, i.e., an extreme (trivial) case, since, by definition, it makes no sense to speak of a SWAP gate if there are not at least two qubits that exchange their states[1].

The ascending order (from top to bottom) corresponds to that presented in all the physical platforms and simulators used in quantum computing for the implementation of algorithms and protocols: Quirk[17], IBM Q Experience[18], Rigetti[19], Quantum Programming Studio[20], Quantum Inspire by QuTech[21], and so on. As it will be explained in the next section, this arrangement will show different tools (in particular, the density matrix) in an opposite way to the original mathematical formulation (matrix treatment), which must be taken into account to correctly understand the results of the experiments that will be obtained in later sections.

Next, and without loss of generality, the four mentioned degrees of QFG will be exposed for some conspicuous cases regarding the number of qubits involved (*p*).

*Zero-degree QFG:* This gate leaves the inputs as is regardless of the number of *q* qubits involved, being zero the number of QFT blocks used.

For a generic number of qubits *p*, QFG will be,

$$F_p^0 = SWAP_{2^p \times 2^p} \; QFT_{2^p \times 2^p}^0 \; SWAP_{2^p \times 2^p} = SWAP_{2^p \times 2^p} \; SWAP_{2^p \times 2^p} = I_{2^p \times 2^p}, \quad \forall p. \tag{8}$$

Specifically, for *p* = 2, the result is found in Eq.(4) and that is repeated in Eq.(9) with the absence of QFT blocks,

$$F_2^0 = SWAP_{2^2 \times 2^2} \; SWAP_{2^2 \times 2^2} = I_{2^2 \times 2^2}. \tag{9}$$

The importance of this gate will be appreciated in the next section.
*First-degree QFG:* This gate implies a single QFT block between both SWAP gates,

$$F_p^1 = SWAP_{2^p \times 2^p} \; QFT_{2^p \times 2^p}^1 \; SWAP_{2^p \times 2^p}, \quad \forall p, \tag{10}$$

where Eq.(10) constitutes a key piece of the teleportation protocol without entanglement that will be developed in a later section. Moreover, a relevant example of QFG for a single qubit is represented by the Hadamard matrix[3-6],

$$F_1^1 = SWAP_{2^1 \times 2^1} \; QFT_{2^1 \times 2^1}^1 \; SWAP_{2^1 \times 2^1} = I_{2^1 \times 2^1} \; H \; I_{2^1 \times 2^1} = H, \tag{11}$$

being $QFT_{2^1 \times 2^1} = H$, i.e., the first-degree version of the QFG for a single qubit is the same Hadamard matrix[3-6], since, as shown in Eq.(7), any application of the SWAP gate on any other gate for the case of a single qubit results in the same gate, since according to Eq.(7) it is the identity matrix.
*Second-degree QFG:* This gate involves two QFT blocks between both SWAP gates, and as established in previous works[3-6], it constitutes a fundamental piece in the generation of entanglement[1], which will be extensively exploited in a later section. For the case of a generic number of qubits *p*, the superscript 2 means 2 QFT modules (matrices of $2^p \times 2^p$ elements each one), the gate is expressed as,

$$F_p^2 = SWAP_{2^p \times 2^p} \; QFT_{2^p \times 2^p}^2 \; SWAP_{2^p \times 2^p}, \quad \forall p. \tag{12}$$

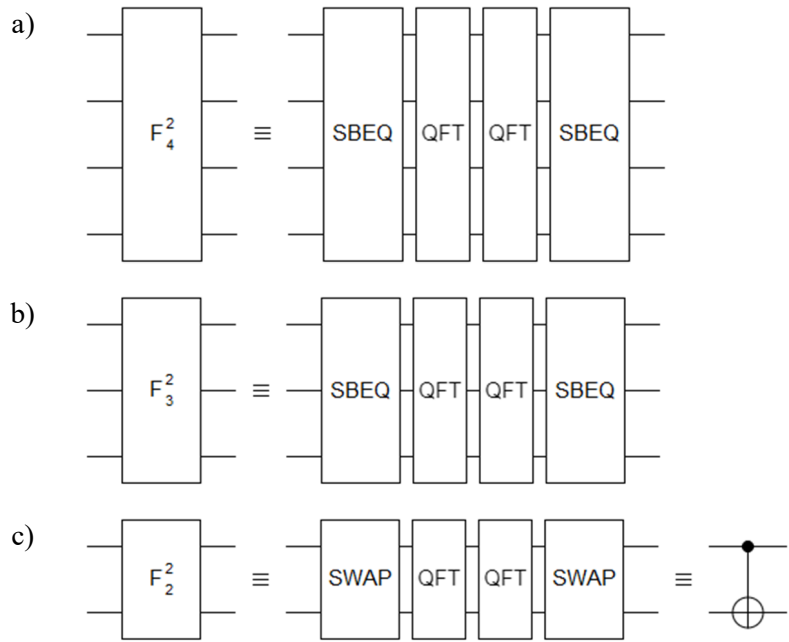

FIG. 2. Second-degree QFG for a) 4 qubits, b) 3 qubits, and c) 2 qubits, where SBEQ = SWAP gate.

Next, examples of this gate are developed for the 4, 3, 2, and 1 qubit cases. The first case is for 4 qubits, resulting in the following gate,

$$F_4^2 = SWAP_{2^4 \times 2^4} \; QFT_{2^4 \times 2^4}^2 \; SWAP_{2^4 \times 2^4}, \tag{13}$$

which can be seen in Fig. 2(a). This gate will be essential in the implementation of the $GHZ_4$ state. The second case is for 3 qubits, where the gate of Fig. 2(b) is obtained,

$$F_3^2 = SWAP_{2^3 \times 2^3} \; QFT_{2^3 \times 2^3}^2 \; SWAP_{2^3 \times 2^3}. \tag{14}$$

This gate will be essential in the implementation of the $GHZ_3$ state. The third example has to do with Fig. 2(c), where the gate resulting ends up being the Feynman's gate,

$$F_2^2 = SWAP_{2^2 \times 2^2} \; QFT_{2^2 \times 2^2}^2 \; SWAP_{2^2 \times 2^2} = CNOT. \tag{15}$$

This case was extensively treated in previous works[3-6], and as already mentioned, it is of wide application in both entanglement[1] and quantum teleportation[8].

Finally, taking into account Eq.(7), and considering that multiplication of two Hadamard[1] matrices $H \in \mathbb{C}^{2^1 \times 2^1}$ is the identity matrix,

$$H\,H = \frac{1}{\sqrt{2}} \begin{bmatrix} 1 & 1 \\ 1 & -1 \end{bmatrix} \frac{1}{\sqrt{2}} \begin{bmatrix} 1 & 1 \\ 1 & -1 \end{bmatrix} = \begin{bmatrix} 1 & 0 \\ 0 & 1 \end{bmatrix} = I_{2^1 \times 2^1}, \tag{16}$$

the corresponding QFG results,

$$F_1^2 = SWAP_{2^1 \times 2^1} \; QFT_{2^1 \times 2^1}^2 \; SWAP_{2^1 \times 2^1}. \tag{17}$$

Remembering that[3-6] $QFT_{2^1 \times 2^1} = H$, and taking into account Eq. (16), Eq. (17) is redefined as,

$$\begin{aligned} F_1^2 &= SWAP_{2^1 \times 2^1} \; QFT_{2^1 \times 2^1}^2 \; SWAP_{2^1 \times 2^1} = SWAP_{2^1 \times 2^1} \; QFT_{2^1 \times 2^1} \; QFT_{2^1 \times 2^1} \; SWAP_{2^1 \times 2^1} \\ &= I_{2^1 \times 2^1} \; H\,H \; I_{2^1 \times 2^1} = I_{2^1 \times 2^1} I_{2^1 \times 2^1} = I_{2^1 \times 2^1}. \end{aligned} \tag{18}$$

That is, Eq.(18) shows that $F_1^2$ is the identity matrix, which will be used in the section corresponding to quantum teleportation[1].

*Comparison between Toffoli ($T_p$) and $F_p^2$ gates:* Next, a presentation by the opposition will take place between both gates to establish the main logical differences between them, that is, between $F_p^2$ and a known gate like Toffoli[1] for the case in which the inputs are computational basis states[1] (CBS) $\left\{ |0\rangle = \begin{bmatrix} 1 \\ 0 \end{bmatrix}, |1\rangle = \begin{bmatrix} 0 \\ 1 \end{bmatrix} \right\}$, where $|0\rangle$ and $|1\rangle$ are the north and south poles of the Bloch sphere[1], respectively.

The aforementioned comparison will take place concerning Fig. 3 for which (∧, ∨, $\veebar$) represent the logical gates AND (logical conjunction), OR (logical disjunction), and XOR (exclusive OR, that is, A$\veebar$B=$\overline{A}\wedge B \vee A \wedge \overline{B}$), respectively.

Figure 3(a) shows the Toffoli gate for 4 qubits where the q[3] output is $q[3]\veebar(q[0]\wedge q[1]\wedge q[2])$ while in the case of the other outputs they are equal to the respective inputs. However, Fig. 3(b) shows the logical outputs of $F_4^2$, where the respective outputs are: q[0], q[0]$\veebar$q[1], $q[2]\veebar(q[0]\vee q[1])$, and $q[3]\veebar(q[0]\vee q[1]\vee q[2])$. Figure 3(c and d) shows both cases for 3 qubits.

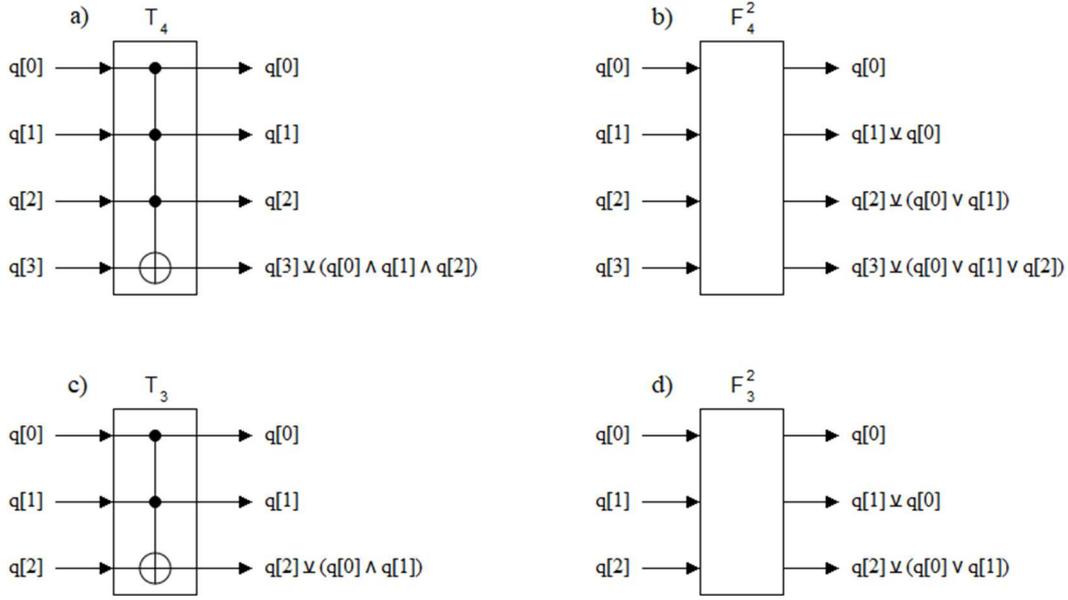

FIG. 3. Toffoli vs QFG of second degree: a) Toffoli of 4 qubits, b) QFG of 4 qubits, c) Toffoli of 3 qubits, and QFG of 3 qubits.

In a general case, we will have,

Toffoli ($T_{i+1}$):

$$q[k] = q[k], \forall k \in [0, i-1], \text{ and} \tag{19a}$$

$$q[i] = q[i] \veebar \left( \bigwedge_{j=0}^{i-1} q[j] \right), \text{ for } k = i, \text{ and} \tag{19b}$$

QFG ($F_{i+1}^2$):

$$q[0] = q[0], \text{ and} \tag{20a}$$

$$q[i] = q[i] \veebar \left( \bigvee_{j=0}^{i-1} q[j] \right), \forall k \in [1, i]. \tag{20b}$$

*Third-degree QFG:* At first, this gate returns complementary results to those obtained with the first-degree gate seen above. This will become apparent in the next section. It implies the use of three QFT blocks in the middle of both SWAP gates,

$$F_p^3 = SWAP_{2^p \times 2^p} \, QFT_{2^p \times 2^p}^3 \, SWAP_{2^p \times 2^p}, \quad \forall p. \tag{21}$$

For degrees greater than 3, the following equivalences arise:

$$F_p^4 = SWAP_{2^p \times 2^p} \, QFT_{2^p \times 2^p}^4 \, SWAP_{2^p \times 2^p} = F_p^0, \quad \forall p, \tag{22}$$

and in this way coincidences between gates that differ by 4 degrees begin to manifest independently of the number of qubits $p$, that is,

$$F_p^0 = F_p^4 = F_p^8 = \cdots = F_p^{0+k4}, \tag{23a}$$

$$F_p^1 = F_p^5 = F_p^9 = \cdots = F_p^{1+k4}, \tag{23b}$$

$$F_p^2 = F_p^6 = F_p^{10} = \cdots = F_p^{2+k4}, \text{ and} \tag{23c}$$

$$F_p^3 = F_p^7 = F_p^{11} = \cdots = F_p^{3+k4}. \tag{23d}$$

In general, it turns out that

$$F_p^d = F_p^{d+k4}, \tag{24}$$

where, in all cases, "the number of QFT blocks" - $d = k \times 4 \;/\; k \in \mathbb{Z}$, however, this is valid only for values of $k \geq 0$, i.e., $k \in \mathbb{N}_0$ (natural with zero).

***Quantum Fourier states***.- In the same way that there are four families of quantum Fourier gates (QFG), there are four families of quantum Fourier states (QFS). This is because QFGs is the central engine in the generation of QFSs. In general terms, the QFS will depend on the degree of the QFG used and this in turn on the number of qubits involved, although for the particular case of dealing with two qubits, that dependence will extend to the type of CBS that will constitute the input qubits of each configuration, i.e., spin-up $|0\rangle$ or spin-down $|1\rangle$. Therefore, since these states directly depend on their corresponding QFG, there will only be four possible degrees for them.

From now on, and as a consequence of the nomenclature adopted for the QGS, we will generically represent the QFS as follows: $|F\rangle_p^d$, where $d$ is the degree of the QFG gate, i.e., the number of QFT blocks it contains, while $p$ is the number of qubits involved.

Finally, the gates of Eqs. (11, and 18), and their respective equivalences will be fundamental in obtaining the preliminary conclusions that we will arrive at at the end of this section.

***Zero-degree QFS:*** In the generic case of $p$ qubits, for all inputs equal to $|0\rangle$, and considering Eq.(8) and that $\otimes$ is the Kronecker's product[1], it results,

$$|F\rangle_p^0 = \left(H \otimes I_{2^{p-1} \times 2^{p-1}}\right)|0\rangle^{\otimes p} = I_{2^p \times 2^p}\left(H \otimes I_{2^{p-1} \times 2^{p-1}}\right)|0\rangle^{\otimes p} = F_p^0\left(F_1^1 \otimes F_{p-1}^0\right)|0\rangle^{\otimes p} = |+\rangle|0\rangle^{\otimes p-1}, \tag{25}$$

where $\left\{|+\rangle = H|0\rangle = \begin{bmatrix} 1/\sqrt{2} \\ 1/\sqrt{2} \end{bmatrix}, |-\rangle = H|1\rangle = \begin{bmatrix} 1/\sqrt{2} \\ -1/\sqrt{2} \end{bmatrix}\right\}$. While in the particular case of two qubits, and considering Eq. (9), with $|F\rangle_1^2 = |F\rangle_1^0$, we will have four states $|F_{ab}\rangle_2^0$, where $a$ is the *phase bit* and $b$ is the *parity bit*. In consequence, these four states $|F_{ab}\rangle_2^0$, in terms of phase and parity bits, will be,

$$|X^a +\rangle|b\rangle = \left(H \otimes I_{2^1 \times 2^1}\right)|ab\rangle = I_{2^2 \times 2^2}\left(H \otimes I_{2^1 \times 2^1}\right)|ab\rangle = F_2^0\left(F_1^1 \otimes F_1^2\right)|ab\rangle = |F_{ab}\rangle_2^0, \tag{26}$$

where $X = \begin{bmatrix} 0 & 1 \\ 1 & 0 \end{bmatrix}$ is the inverter gate[1]. Table 1 shows the $|F_{ab}\rangle_2^0$ states in terms of phase and parity bits, while Fig. 4 represents the $|F_{ab}\rangle_2^0$ states thanks to three different representations, being the last one implemented exclusively in terms of QFGs. Finally, Table 2 shows the density matrix of the four $|F_{ab}\rangle_2^0$ states in terms of phase and parity bits, where the density matrix is equal to $|F_{ab}\rangle_2^0 \langle F_{ab}|_2^0$.

| $a$ | $b$ | $\lvert F_{ab}\rangle_2^0$ |
|---|---|---|
| 0 | 0 | $\lvert +0\rangle$ |
| 1 | 0 | $\lvert -0\rangle$ |
| 0 | 1 | $\lvert +1\rangle$ |
| 1 | 1 | $\lvert -1\rangle$ |

Table 1. $\lvert F_{ab}\rangle_2^0$ states in terms of phase ($a$) and parity ($b$) bits.

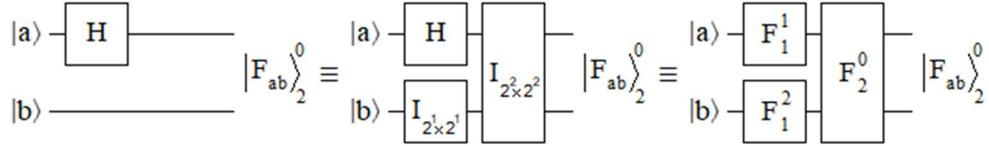

FIG. 4. $\lvert F_{ab}\rangle_2^0$ states via different representations, where the last one is exclusively in terms of QFGs.

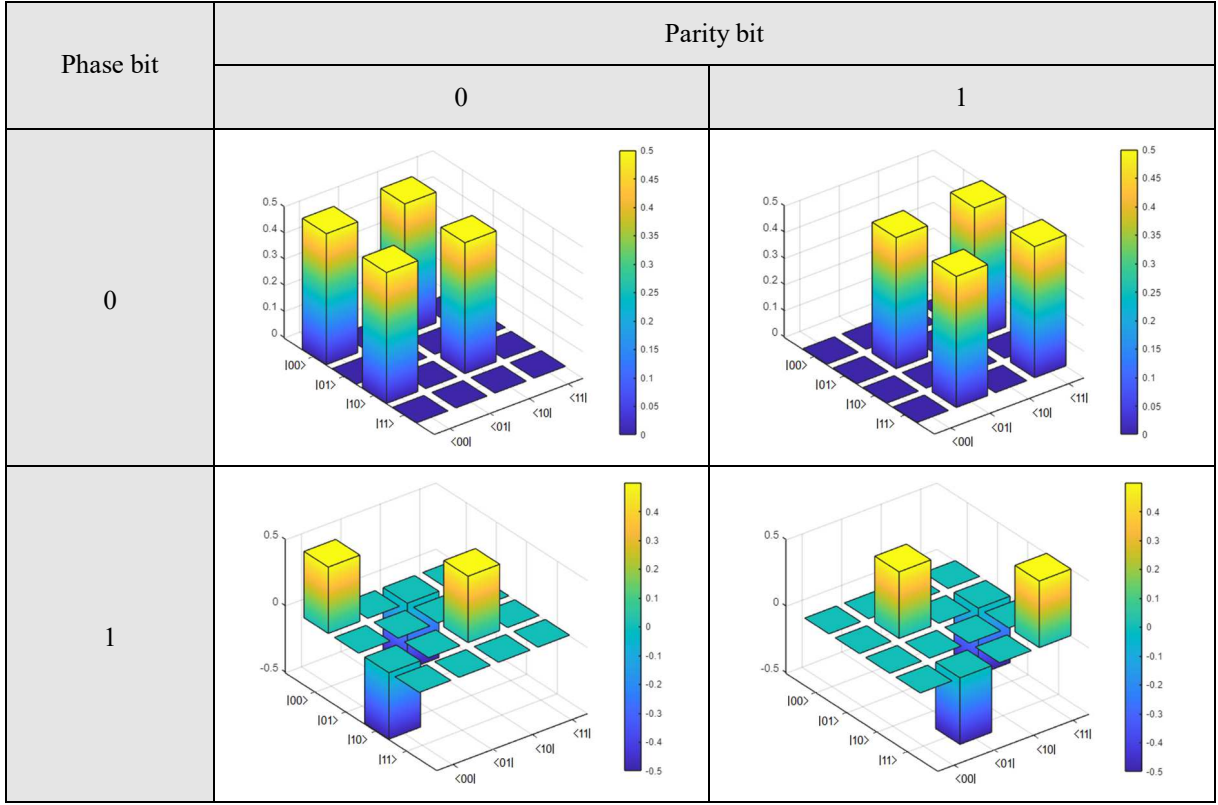

Table 2. $\lvert F_{ab}\rangle_2^0$ states density matrices in terms of phase and parity bits.

*First-degree QFS:* In the generic case of $p$ qubits, and for all its inputs equal to $\lvert 0\rangle$, it turns out

$$\lvert F\rangle_p^1 = F_p^1\left(H\otimes I_{2^{p-1}\times 2^{p-1}}\right)\lvert 0\rangle^{\otimes p} = F_p^1\left(F_1^1\otimes F_{p-1}^0\right)\lvert 0\rangle^{\otimes p}. \tag{27}$$

In the particular case of two qubits, we will have

$$|F_{ab}\rangle_2^1 = F_2^1\left(H \otimes I_{2^1 \times 2^1}\right)|ab\rangle = F_2^1\left(F_1^1 \otimes F_1^2\right)|ab\rangle$$
$$= \begin{bmatrix} \bar{a} & a & (-i)^a(-1)^b(1+i)/2 & (i)^a(-1)^b(1-i)/2 \end{bmatrix}^T / \sqrt{2} \qquad (28)$$

where $\bar{a}$ is the inverse of $a$, i.e., if $a = 0$, then, $\bar{a} = 1$, and vice versa, $(\bullet)^T$ means *transpose of* $(\bullet)$, and $i = \sqrt{-1}$. Figure 5 represents the four $|F_{ab}\rangle_2^1$ states thanks to two different representations, being the last one implemented exclusively in terms of QFGs, while Table 3 shows the $|F_{ab}\rangle_2^1$ states in terms of phase and parity bits.

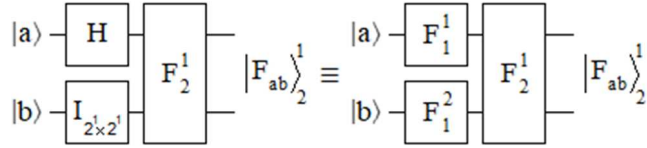

FIG. 5. $|F_{ab}\rangle_2^1$ states via two representations, where the last one is exclusively in terms of QFGs.

| $a$ | $b$ | $|F_{ab}\rangle_2^1$ |
|---|---|---|
| 0 | 0 | $\begin{bmatrix} 1 & 0 & (1+i)/2 & (1-i)/2 \end{bmatrix}^T / \sqrt{2}$ |
| 1 | 0 | $\begin{bmatrix} 0 & 1 & (1-i)/2 & (1+i)/2 \end{bmatrix}^T / \sqrt{2}$ |
| 0 | 1 | $\begin{bmatrix} 1 & 0 & -(1+i)/2 & -(1-i)/2 \end{bmatrix}^T / \sqrt{2}$ |
| 1 | 1 | $\begin{bmatrix} 0 & 1 & -(1-i)/2 & -(1+i)/2 \end{bmatrix}^T / \sqrt{2}$ |

Table 3. $|F_{ab}\rangle_2^1$ states in terms of phase ($a$) and parity ($b$) bits.

As we can see both in Eq. (28) and in Table 3, these states do not constitute a maximally-entangled pair[1] or even a non-maximally-entangled pair[22-25]. The density matrices for the four cases of $|F_{ab}\rangle_2^1$ (that is, according to the phase and parity bits) can be seen in Table 4, where the complexity increases concerning the case of $|F_{ab}\rangle_2^0$ because they are complex matrices with real and imaginary parts.

The density matrices arise from the external products $|F_{ab}\rangle_2^1 \langle F_{ab}|_2^1$, and they are a fundamental witness element that confirms that it is not about maximally or non-maximally entangled states. This is reflected by the elements that are occupied in those arrays, where even in the case of non-maximally-entangled states[1], the elements occupied are the same as in the case of maximally-entangled states[22-25]. This will be seen in detail in the subsection called *Preliminary Conclusions*, at the end of this section.

As we will see in the next section, even if it is not any form of known entanglement, the $|F_{ab}\rangle_2^1$ states will give rise to valid forms of teleportation[8] where this characteristic (that is, not being some traditional form of entanglement) will not condition its performance at all, when these states are used in the context of quantum communications[7], particularly in the future quantum Internet[9-14].

*Second-degree QFS:* In the generic case of $p$ qubits, and for all its inputs equal to $|0\rangle$, it turns out

$$|F\rangle_p^2 = F_p^2\left(H \otimes I_{2^{p-1} \times 2^{p-1}}\right)|0\rangle^{\otimes p} = F_p^2\left(F_1^1 \otimes F_{p-1}^0\right)|0\rangle^{\otimes p}. \qquad (29)$$

In the particular case of two qubits, and considering Eqs. (3c, and 15), we will have

$$\left|\beta_{ab}\right\rangle = CNOT\left(H \otimes I_{2^1 \times 2^1}\right)\left|ab\right\rangle = F_2^2\left(F_1^1 \otimes F_1^2\right)\left|ab\right\rangle = \left|F_{ab}\right\rangle_2^2 = \left(\left|0b\right\rangle + (-1)^a \left|1, b \veebar 1\right\rangle\right) / \sqrt{2} \quad (30)$$

| Phase bit | Parity bit | Density matrix | |
|---|---|---|---|
| | | real part | imaginary part |
| 0 | 0 | | |
| 0 | 1 | | |
| 1 | 0 | | |
| 1 | 1 | | |

Table 4. $\left|F_{ab}\right\rangle_2^1$ states density matrices in terms of phase and parity bits.

These are the famous Bell states[1], which are a particular case of the Fourier states, that is, the Bell states are the second degree Fourier states. Figure 6 represents the four $\left|F_{ab}\right\rangle_2^2$ states thanks to three different representations, being the last one implemented exclusively in terms of QFGs, while Table 5 shows the $\left|F_{ab}\right\rangle_2^2$ states in terms of phase and parity bits.

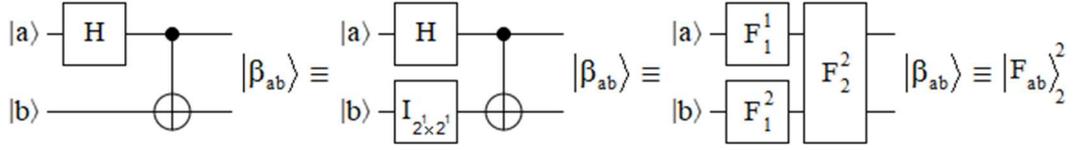

FIG. 6. $|F_{ab}\rangle_2^2$ states via different representations, where the last one is exclusively in terms of QFGs.

| $a$ | $b$ | $|F_{ab}\rangle_2^2$ |
|---|---|---|
| 0 | 0 | $[1\ 0\ 0\ 1]^T/\sqrt{2} = |\beta_{00}\rangle$ |
| 1 | 0 | $[1\ 0\ 0\ -1]^T/\sqrt{2} = |\beta_{10}\rangle$ |
| 0 | 1 | $[0\ 1\ 1\ 0]^T/\sqrt{2} = |\beta_{01}\rangle$ |
| 1 | 1 | $[0\ 1\ -1\ 0]^T/\sqrt{2} = |\beta_{11}\rangle$ |

Table 5. $|F_{ab}\rangle_2^2$ states in terms of phase ($a$) and parity ($b$) bits.

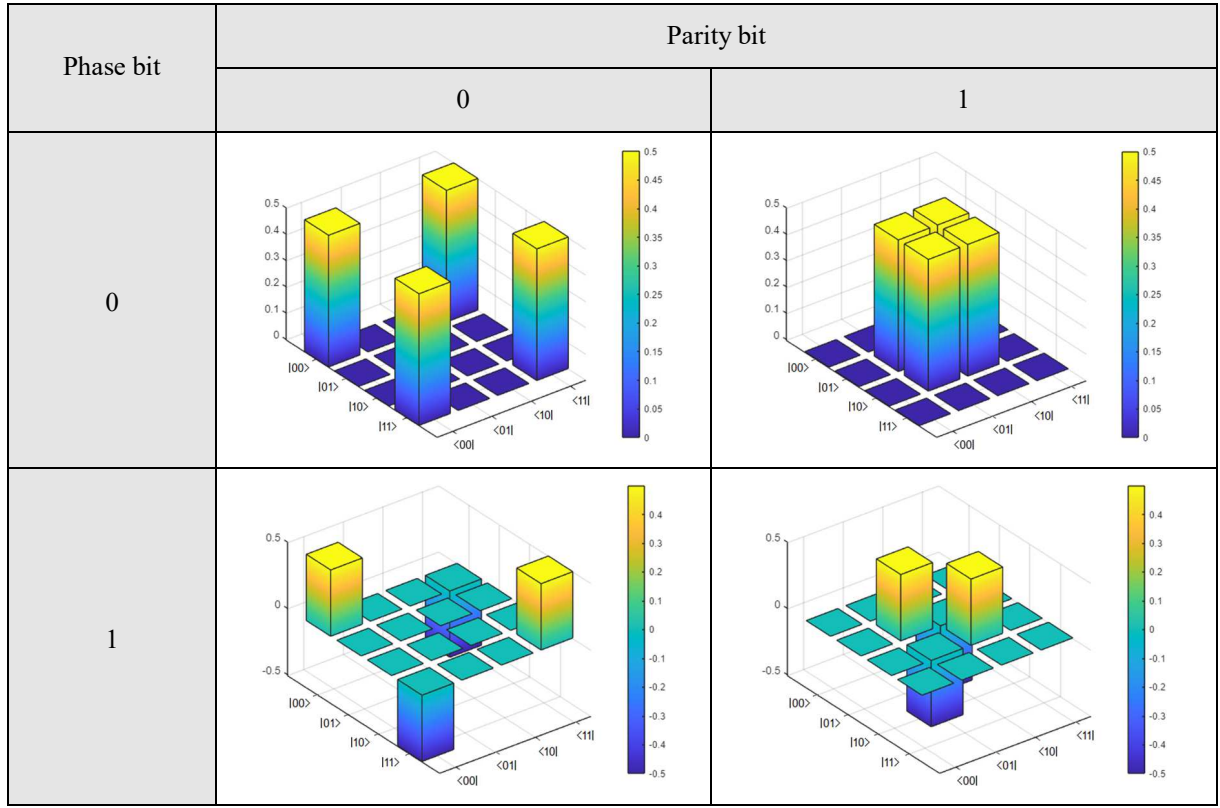

Table 6. $|F_{ab}\rangle_2^2$ states density matrices in terms of phase and parity bits.

The density matrices for the four cases of $|F_{ab}\rangle_2^2$ (that is, according to the phase and parity bits) can be seen in Table 6, which arise from the external products $|F_{ab}\rangle_2^2 \langle F_{ab}|_2^2$. These four matrices represent, like no other tool, the four maximally-entangled states[1]. The positions of the elements occupied in these matrices, as well as the signs of their values, are extremely relevant to be able to identify the type of Bell state to which it refers. These positions are the border that separates the Fourier states of the first degree and the second degree, and the fundamental reason to affirm that the first-degree Fourier states are not even non-maximally entangled states.

As a natural extension of the Bell state $|\beta_{00}\rangle$, we can refer to the Greenberger-Horne-Zeilinger states of 3 and 4 qubits, i.e., $|GHZ_3\rangle$ and $|GHZ_4\rangle$, which are implemented in Figs. 7 and 8, respectively, where the $|GHZ_3\rangle$ is known as $|F_{000}\rangle_3^2$, or $|F\rangle_3^2$, and the three subscript zeros correspond to the three *spin-up* (or $|0\rangle$) inputs of $|GHZ_3\rangle$ configuration.

Figures 7(a) are all completely equivalents, where the last one is implemented exclusively in terms of QFGs, however, this does not mean at all that the nesting of two Feynman gates in the first case of this figure is equivalent to the $F_3^2$ gate. For this case, the same procedure has been followed as in the previous cases, that is, it is a process of approximation to the final equivalence through individual equivalences. Taking into account Eq. (9), $|GHZ_3\rangle$ in terms of QFGs results,

$$|GHZ_3\rangle = [1\ 0\ 0\ 0\ 0\ 0\ 0\ 1]^T/\sqrt{2}$$
$$= (SWAP \otimes I_{2^1 \times 2^1})(I_{2^1 \times 2^1} \otimes CNOT)(SWAP \otimes I_{2^1 \times 2^1})(CNOT \otimes I_{2^1 \times 2^1})(H \otimes I_{2^2 \times 2^2})|000\rangle \quad (31)$$
$$= F_3^2 (F_1^1 \otimes F_2^0)|000\rangle = |F_{000}\rangle_3^2 = |F\rangle_3^2.$$

Finally, Fig. 7(b) represents the density matrix of $|F\rangle_3^2$ state, that is $|F\rangle_3^2 \langle F|_3^2$, which is very similar to that of the Bell state $|\beta_{00}\rangle = |F_{00}\rangle_2^2$ of Table 6 though stretched from all four ends.

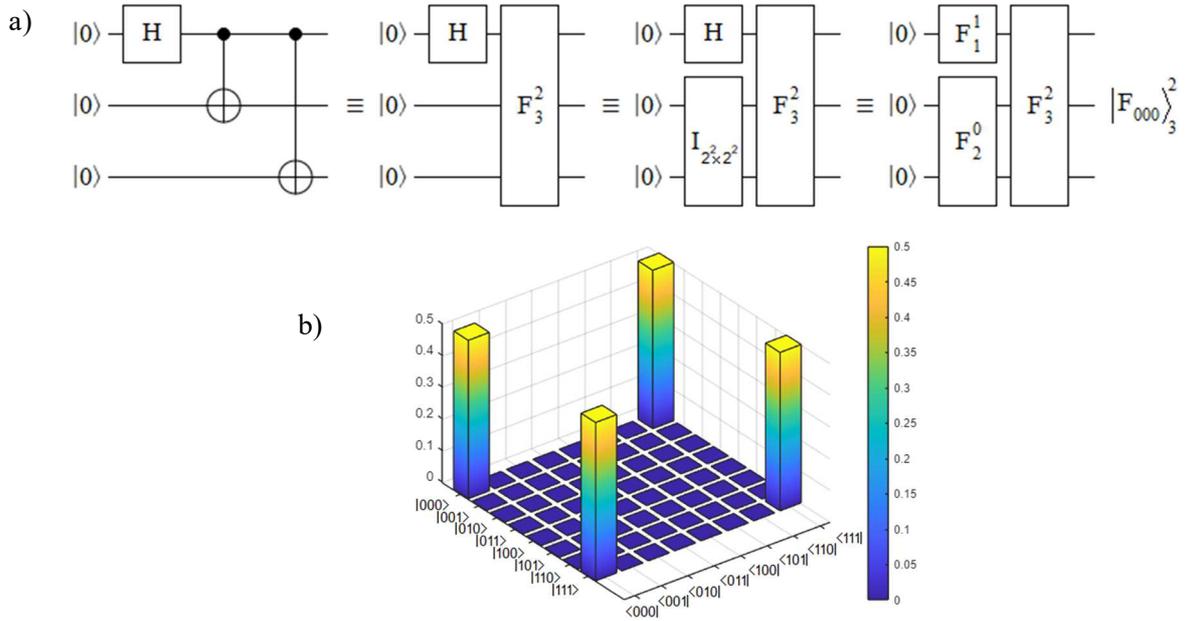

FIG. 7. $|GHZ_3\rangle \equiv |F_{000}\rangle_3^2 = |F\rangle_3^2$ state, where a) represents equivalences between gates, while, b) shows a 3D implementation of its density matrix.

Figures 8(a) shows four equivalences of $|GHZ_4\rangle$, where the last one is implemented exclusively in terms of QFGs, while Fig. 8(b) represents the density matrix of the $|F\rangle_4^2$ state, that is $|F\rangle_4^2 \langle F|_4^2$, with the same similarity considerations as those expressed in the previous case regarding the density matrix of the Bell state $|\beta_{00}\rangle$. For this particular case, that is, $|GHZ_4\rangle \equiv |F_{0000}\rangle_4^2 = |F\rangle_4^2$, and considering that $F_3^0 = I_{2^3 \times 2^3}$, we will have,

$$|GHZ_4\rangle = [1\ 0\ 0\ 0\ 0\ 0\ 0\ 0\ 0\ 0\ 0\ 0\ 0\ 0\ 0\ 1]^T/\sqrt{2}$$
$$= (SWAP \otimes I_{2^2 \times 2^2})(I_{2^1 \times 2^1} \otimes SWAP \otimes I_{2^1 \times 2^1})(I_{2^2 \times 2^2} \otimes CNOT)(I_{2^1 \times 2^1} \otimes SWAP \otimes I_{2^1 \times 2^1})(SWAP \otimes I_{2^2 \times 2^2}) \quad (32)$$
$$(SWAP \otimes I_{2^2 \times 2^2})(I_{2^1 \times 2^1} \otimes CNOT \otimes I_{2^1 \times 2^1})(SWAP \otimes I_{2^2 \times 2^2})(CNOT \otimes I_{2^2 \times 2^2})(H \otimes I_{2^3 \times 2^3})|0000\rangle$$
$$= F_4^2 (F_1^1 \otimes F_3^0)|0000\rangle = |F_{0000}\rangle_4^2 = |F\rangle_4^2.$$

Finally, at this point, similar considerations are made to those clarified for the case of $|GHZ_3\rangle$ in that the nesting of *CNOT* gates at the beginning of Fig. 8(a) is not equivalent to the $F_4^2$ gate, although all implementations of this figure are, i.e., they all have output to $|GHZ_4\rangle \equiv |F_{0000}\rangle_4^2 = |F\rangle_4^2$.

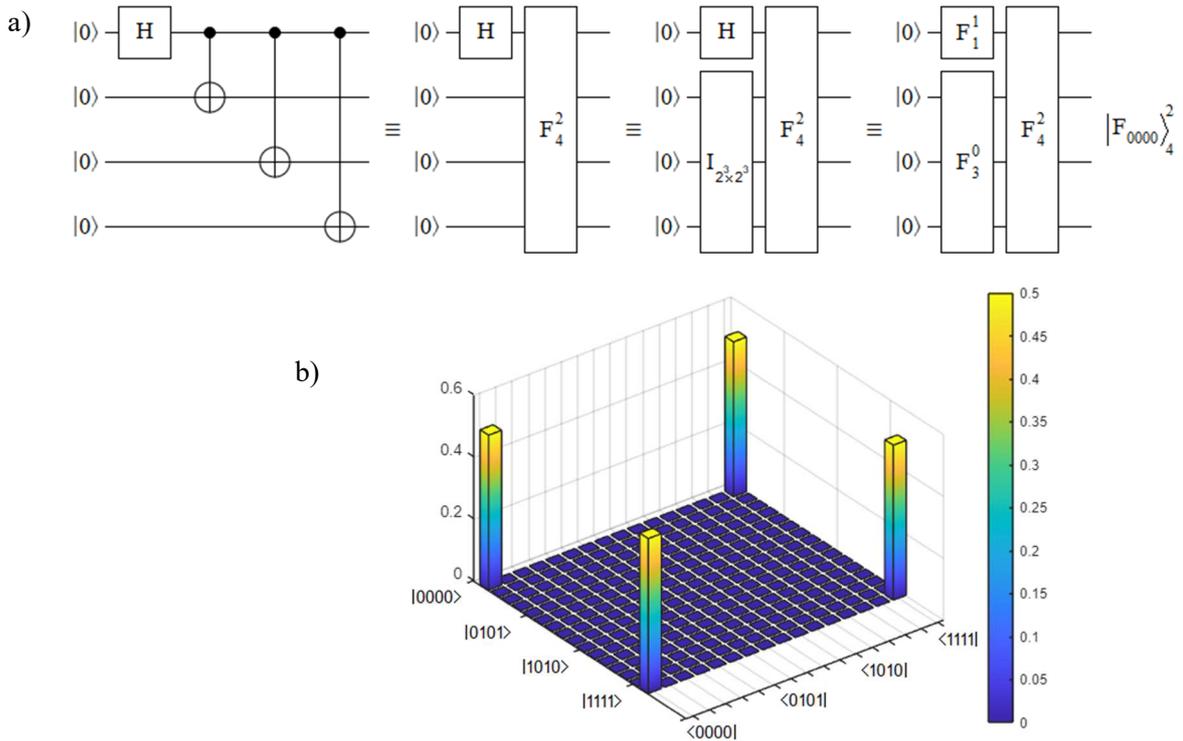

FIG. 8. $|GHZ_4\rangle \equiv |F_{0000}\rangle_4^2 = |F\rangle_4^2$ state, where a) represents equivalences between gates, while, b) shows a 3D implementation of its density matrix.

*Third-degree QFS:* In the generic case of *p* qubits, for all its inputs equal to $|0\rangle$, and taking into account Eq. (8), it turns out

$$|F\rangle_p^3 = F_p^3 (H \otimes I_{2^{p-1} \times 2^{p-1}})|0\rangle^{\otimes p} = F_p^3 (F_1^1 \otimes F_{p-1}^0)|0\rangle^{\otimes p}. \quad (33)$$

In the particular case of two qubits, we will have

$$|F_{ab}\rangle_2^3 = F_2^3 (H \otimes I_{2^1 \times 2^1})|ab\rangle = F_2^3 (F_1^1 \otimes F_1^2)|ab\rangle$$
$$= [\bar{a}\ a\ (i)^a (-1)^b (1-i)/2\ (-i)^a (-1)^b (1+i)/2]^T / \sqrt{2}. \quad (34)$$

Figure 9 represents the four $|F_{ab}\rangle_2^3$ states thanks to two different representations, being the last one that implemented exclusively in terms of QFGs, while Table 7 shows the $|F_{ab}\rangle_2^3$ states in terms of phase and parity bits. As in the case of Tables 3 and 4 for $|F_{ab}\rangle_2^1$, Table 7 shows that these states also do not constitute a maximally-entangled pair[1] or even a non-maximally-entangled pair[22-25]. The density matrices for the four cases of $|F_{ab}\rangle_2^3$ (that is, according to the phase and parity bits) can be seen in Table 8, where their complexity is similar to that of state $|F_{ab}\rangle_2^1$, because they are complex matrices with real and imaginary parts. This density matrix is $|F_{ab}\rangle_2^3 \langle F_{ab}|_2^3$ and its elements occupy practically the same positions as those of the density matrix of the state $|F_{ab}\rangle_2^1$. Moreover, as can be seen, the third-degree states $|F_{ab}\rangle_2^3$ are complex conjugates of those of the first-degree states $|F_{ab}\rangle_2^1$.

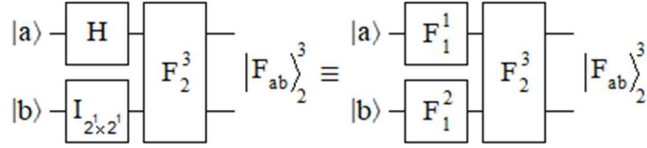

FIG. 9. $|F_{ab}\rangle_2^3$ states via two representations, where the last one is exclusively in terms of QFGs.

| $a$ | $b$ | $|F_{ab}\rangle_2^3$ |
|---|---|---|
| 0 | 0 | $\begin{bmatrix} 1 & 0 & (1-i)/2 & (1+i)/2 \end{bmatrix}^T/\sqrt{2}$ |
| 1 | 0 | $\begin{bmatrix} 0 & 1 & (1+i)/2 & (1-i)/2 \end{bmatrix}^T/\sqrt{2}$ |
| 0 | 1 | $\begin{bmatrix} 1 & 0 & -(1-i)/2 & -(1+i)/2 \end{bmatrix}^T/\sqrt{2}$ |
| 1 | 1 | $\begin{bmatrix} 0 & 1 & -(1+i)/2 & -(1-i)/2 \end{bmatrix}^T/\sqrt{2}$ |

Table 7. $|F_{ab}\rangle_2^3$ states in terms of phase ($a$) and parity ($b$) bits.

*Non-maximally entangled states:* There are several versions to represent non-maximally entangled states[22-25], so we will choose one[22], by which we will replace the Hadamard matrix with another gate, for example,

$$\sqrt[4]{X} = \sqrt[4]{\begin{bmatrix} 0 & 1 \\ 1 & 0 \end{bmatrix}} = \begin{bmatrix} 0.8536+0.3536i & 0.1464-0.3536i \\ 0.1464-0.3536i & 0.8536+0.3536i \end{bmatrix} = \begin{bmatrix} u & v \\ v & u \end{bmatrix}, \quad (35)$$

where $u = 0.8536+0.3536i$, and $v = 0.1464-0.3536i$. Based on Fig. 10, the resulting state of this two-qubit configuration for the non-maximally entangled case will be,

$$|\gamma_{ab}\rangle = CNOT\left(\sqrt[4]{X} \otimes I_{2^1 \times 2^1}\right)|ab\rangle. \quad (36)$$

The two first implementations of Fig. 10 show Eq. (36) depending on the phase and parity bits, while Table 9 shows the four resulting states. The corresponding four density matrices of the states $|\gamma_{ab}\rangle$ are $|\gamma_{ab}\rangle\langle\gamma_{ab}|$, and they can be seen in Fig. 11. The positions that the elements occupy in the density matrices are similar to the maximally-entangled case, only the values change according to the matrix chosen to replace the Hadamard matrix ($H$).

| Phase bit | Parity bit | Density matrix ||
|---|---|---|---|
| | | real part | imaginary part |
| 0 | 0 | | |
| 0 | 1 | | |
| 1 | 0 | | |
| 1 | 1 | | |

Table 8. $\left|F_{ab}\right\rangle_2^3$ states density matrices in terms of phase and parity bits.

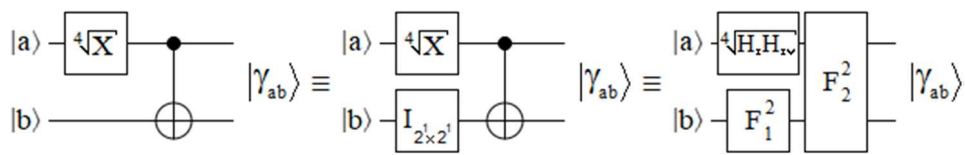

FIG. 10. $\left|\gamma_{ab}\right\rangle$ states via three representations, where the last one is exclusively in terms of QFGs.

Pauli's matrices[1] can be expressed in terms of the so-named Hadamard rotation gates[26] or the general unitary operator $U(\theta,\varphi,\lambda) = \begin{bmatrix} cos(\theta/2) & -e^{i\lambda}sin(\theta/2) \\ e^{i\varphi}sin(\theta/2) & e^{i(\lambda+\varphi)}cos(\theta/2) \end{bmatrix}$ as follows:

$I = H_I H_I = H_{III} H_{III} = H_{II} H_{IV} = H_{IV} H_{II}$, (37a)

$X = H_{III} H_{II} = H_{II} H_I = H_I H_{IV} = H_{IV} H_{III}$, (37b)

$Y = i H_{III} H_I = i H_{II} H_{II} = -i H_{IV} H_{IV} = -i H_I H_{III}$, and (37c)

$Z = -H_{II} H_{III} = H_I H_{II} = -H_{III} H_{IV} = H_{IV} H_I$, (37d)

where $I$ is a 2×2 identity matrix, $i = \sqrt{-1}$, $H_I = H = U(\pi/2, 0, \pi)$ of Eq.(3), while

$H_{II} = \frac{1}{\sqrt{2}}\begin{bmatrix} 1 & -1 \\ 1 & 1 \end{bmatrix} = U(\pi/2, 0, 0)$, $H_{III} = \frac{1}{\sqrt{2}}\begin{bmatrix} -1 & 1 \\ 1 & 1 \end{bmatrix} = U(5\pi/2, \pi, 0)$, and $H_{IV} = \frac{1}{\sqrt{2}}\begin{bmatrix} 1 & 1 \\ -1 & 1 \end{bmatrix} = U(\pi/2, \pi, \pi)$. (38)

Therefore, we rewrite Eq. (36) taking into account Eq. (37b),

$$|\gamma_{ab}\rangle = CNOT\left(\sqrt[4]{X} \otimes I_{2^1 \times 2^1}\right)|ab\rangle = CNOT\left(\sqrt[4]{H_I H_{IV}} \otimes F_I^2\right)|ab\rangle,$$ (39)

where $H_I = F_I^1$, and $H_{IV}$ is $H_I$ after a left-to-right flipping procedure[26].

| $a$ | $b$ | $|\gamma_{ab}\rangle$ |
|---|---|---|
| 0 | 0 | $[u \; 0 \; 0 \; v]^T = |\gamma_{00}\rangle$ |
| 1 | 0 | $[v \; 0 \; 0 \; u]^T = |\gamma_{10}\rangle$ |
| 0 | 1 | $[0 \; u \; v \; 0]^T = |\gamma_{01}\rangle$ |
| 1 | 1 | $[0 \; v \; u \; 0]^T = |\gamma_{11}\rangle$ |

Table 9. $|\gamma_{ab}\rangle$ states in terms of phase ($a$) and parity ($b$) bits.

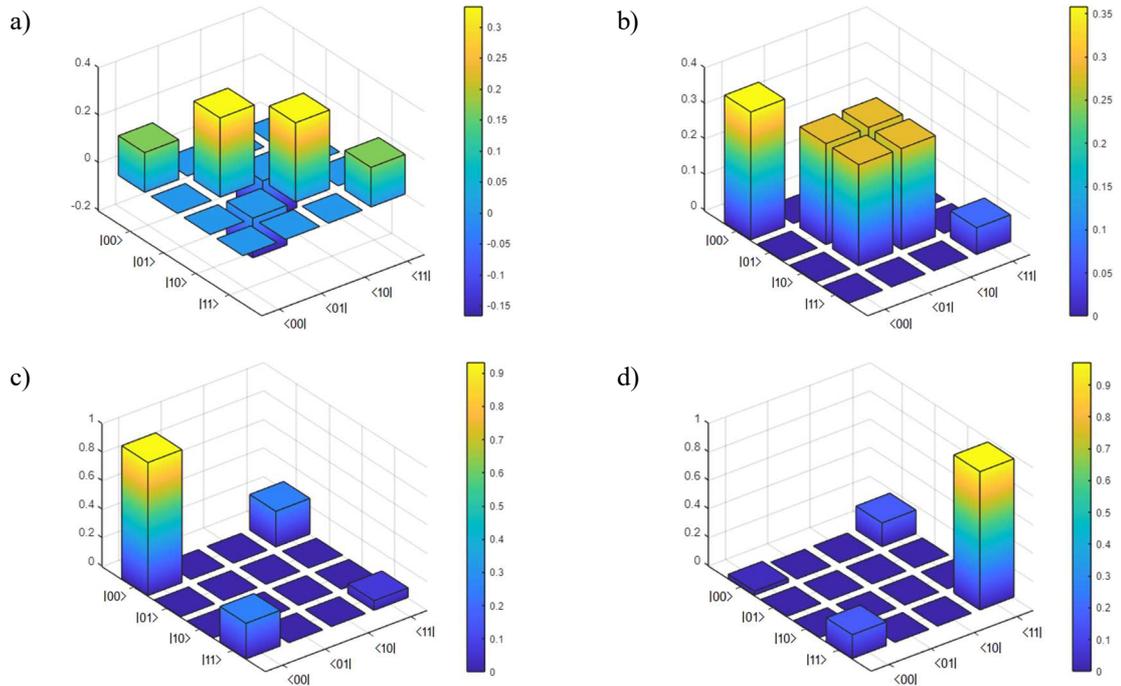

FIG. 11. $|\gamma_{ab}\rangle$ states density matrices for: a) $a = 0, b = 0$, b) $a = 1, b = 0$, c) $a = 0, b = 1$, and d) $a = 1, b = 1$.

*Preliminary conclusions:* As a result of this section, it is possible to arrive at a series of preliminary conclusions about the most outstanding distinctive characteristics of Fourier states:

(1) Entanglement is a particular case of Fourier states, it is just that, one more case,
(2) previous works have shown that entanglement is also a spectral phenomenon[3-6], as well as a temporal one[1],
(3) the density matrices in Tables 4 and 8 (for $|F_{ab}\rangle_2^1$ and $|F_{ab}\rangle_2^3$, respectively) show that they are different entanglements, that is to say, they are not typical entanglements, not even non-maximally entanglements, but rather rough entanglements. This designation is due to it is a rustic, unfinished, incomplete entanglement that is missing a Fourier layer,
(4) Bell states are a special case of Fourier states. Also, the $|GHZ_n\rangle$ states are particular cases of the Fourier states. Both the Bell and $|GHZ_n\rangle$ states can be represented by Fourier gates,
(5) the Fourier states preserve the same Modular Arithmetic as for the case of the QFGs, that is, there are coincidences between states that differ by 4 degrees, and this manifests independently of the number of qubits $p$, and it is for this reason that we say that there are only four Fourier states. Then,

$$|F\rangle_p^0 = |F\rangle_p^4 = |F\rangle_p^8 = \cdots = |F\rangle_p^{0+k4}, \tag{40a}$$

$$|F\rangle_p^1 = |F\rangle_p^5 = |F\rangle_p^9 = \cdots = |F\rangle_p^{1+k4}, \tag{40b}$$

$$|F\rangle_p^2 = |F\rangle_p^6 = |F\rangle_p^{10} = \cdots = |F\rangle_p^{2+k4}, \text{ and} \tag{40c}$$

$$|F\rangle_p^3 = |F\rangle_p^7 = |F\rangle_p^{11} = \cdots = |F\rangle_p^{3+k4}. \tag{40d}$$

In general, it turns out that

$$|F\rangle_p^d = |F\rangle_p^{d+k4}, \tag{41}$$

where, in all cases, "number of QFT blocks" - $d = k \times 4 \;/\; k \in \mathbb{Z}$, however, this is valid only for values of $k \geq 0$, i.e., $k \in \mathbb{N}_0$ (natural with zero),

(6) Fourier states of degree 1 are the complex conjugates of those of degree 3 and vice versa,
(7) All configurations of Fig. 7(a) are completely similar in their results, i.e., $|GHZ_3\rangle \equiv |F_{000}\rangle_3^2 = |F\rangle_3^2$. Something similar happens between all the configurations of Figs. 8(a) but for four qubits, which result in $|GHZ_4\rangle \equiv |F_{000}\rangle_4^2 = |F\rangle_4^2$, however, when we exclusively introduce CBS to the inputs of the configurations of Fig. 12 ($|a\rangle$, $|b\rangle$ and $|c\rangle$ for 3 qubits, and $|a\rangle$, $|b\rangle$, $|c\rangle$, and $|d\rangle$ for 4 qubits), the equivalencies break down for some combinations of those inputs.

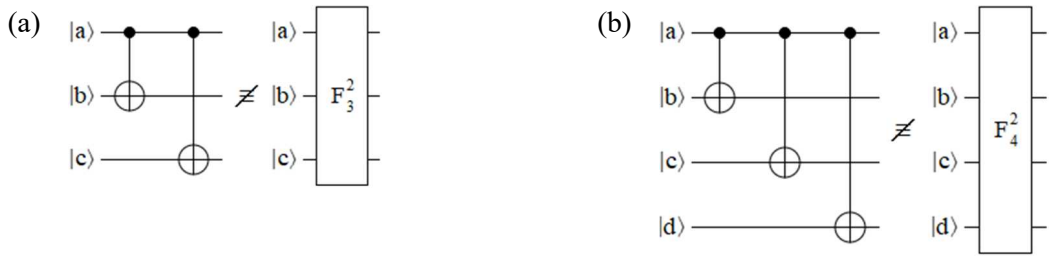

FIG. 12. Equivalences that do not hold between gates according to different combinations of the CBS inputs: a) $|a\rangle$, $|b\rangle$ and $|c\rangle$ for 3 qubits, and b) $|a\rangle$, $|b\rangle$, $|c\rangle$, and $|d\rangle$ for 4 qubits.

An example of these differences for the 3-qubit case is ($|a\rangle = |0\rangle, |b\rangle = |1\rangle$, and $|c\rangle = |0\rangle$) where the outputs are $|011\rangle$ for $F_3^2$ and $|010\rangle$ for the nesting of *CNOTs* in Fig. 12(a), while for the 4-qubit case, one of the differences is ($|a\rangle = |0\rangle, |b\rangle = |1\rangle, |c\rangle = |0\rangle$, and $|d\rangle = |0\rangle$) where the outputs are $|0111\rangle$ for $F_4^2$ and $|0100\rangle$ for the nested *CNOTs* of Fig. 12(b).

(8) The only Fourier states that do not represent some kind of entanglement are those of zero-degree, i.e., without QFT blocks, which shows the correspondence between entanglement and Fourier[3-6].

(9) In the cases of maximally, and non-maximally entangled states, as well as rough entanglement, Fig. 13 represents the positions of non-zero elements in their density matrices as gray tiles. Fig. 13(a) corresponds to $|\beta_{00}\rangle = |F_{00}\rangle_2^2$, $|\beta_{10}\rangle = |F_{10}\rangle_2^2$, $|\gamma_{00}\rangle$, and $|\gamma_{10}\rangle$, Fig. 13(b) represents $|\beta_{01}\rangle = |F_{01}\rangle_2^2$, $|\beta_{11}\rangle = |F_{11}\rangle_2^2$, $|\gamma_{01}\rangle$, and $|\gamma_{11}\rangle$, Fig. 13(c) contains the gray tiles of $|F_{00}\rangle_2^1$, $|F_{01}\rangle_2^1$, $|F_{00}\rangle_2^3$ and $|F_{01}\rangle_2^3$, while Fig. 13(d) shows $|F_{10}\rangle_2^1$, $|F_{11}\rangle_2^1$, $|F_{10}\rangle_2^3$, and $|F_{11}\rangle_2^3$. The same locations are occupied by nonzero elements for the equivalent cases of maximally, and non-maximally entangled states, i.e., for the same combination of phase and parity bits. However, for the case of rough entanglement (for both $|F_{ab}\rangle_2^1$ and $|F_{ab}\rangle_2^3$) the positions occupied by non-zero elements in their density matrices are completely different (regardless of whether some elements are complex), to those of Figs. 13(a, and b). The marked difference between the tiles occupied by the cases maximally entangled states and non-maximally entangled states on one hand, and rough entanglement on the other hand indicates that we are in the presence of another entanglement type. A simple visual inspection of Figs. 13(a, and b) and 13(c, and d) tells us that rough entanglement is a very different case from previously known entanglements[1]. However, and as we will see in the next section, this rustic form of entanglement will allow the successful teleportation of various types of qubits.

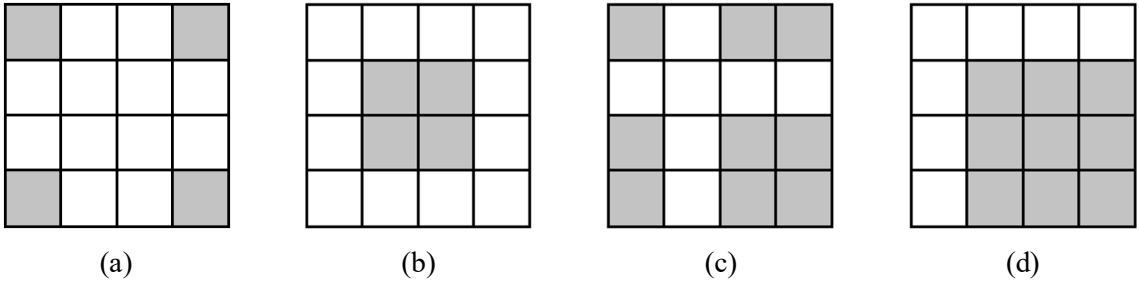

FIG. 13. Gray tiles indicating the non-zero elements of the respective density matrices for a) $|\beta_{00}\rangle = |F_{00}\rangle_2^2$, $|\beta_{10}\rangle = |F_{10}\rangle_2^2$, $|\gamma_{00}\rangle$, and $|\gamma_{10}\rangle$, b) $|\beta_{01}\rangle = |F_{01}\rangle_2^2$, $|\beta_{11}\rangle = |F_{11}\rangle_2^2$, $|\gamma_{01}\rangle$, and $|\gamma_{11}\rangle$, c) $|F_{00}\rangle_2^1$, $|F_{01}\rangle_2^1$, $|F_{00}\rangle_2^3$ and $|F_{01}\rangle_2^3$, and d) $|F_{10}\rangle_2^1$, $|F_{11}\rangle_2^1$, $|F_{10}\rangle_2^3$, and $|F_{11}\rangle_2^3$.

A few examples of the respective density matrices highlight everything previously expressed, that is, based on Eq. (30) we have a pair of cases of maximally-entangled states, such that for $|a\rangle = |0\rangle$ and, $|b\rangle = |0\rangle$, the resulting Bell state is:

$$|\beta_{00}\rangle = (|00\rangle + |11\rangle)/\sqrt{2}$$
$$= \begin{bmatrix} 1 & 0 & 0 & 1 \end{bmatrix}^T / \sqrt{2},$$
(42)

and its corresponding density matrix is,

$$|\beta_{00}\rangle\langle\beta_{00}| = (|00\rangle+|11\rangle)/\sqrt{2}\,(\langle 00|+\langle 11|)/\sqrt{2}$$
$$= (|00\rangle\langle 00|+|11\rangle\langle 00|+|00\rangle\langle 11|+|11\rangle\langle 11|)/2 \qquad (43)$$
$$= \frac{1}{2}\begin{bmatrix} 1 & 0 & 0 & 1 \\ 0 & 0 & 0 & 0 \\ 0 & 0 & 0 & 0 \\ 1 & 0 & 0 & 1 \end{bmatrix}.$$

This matrix represents one of the most conspicuous examples that respond to the characteristics of Fig. 13(a), while for $|a\rangle=|0\rangle$ and $|b\rangle=|1\rangle$, the corresponding Bell state turns out to be,

$$|\beta_{01}\rangle = (|01\rangle+|10\rangle)/\sqrt{2}$$
$$= \begin{bmatrix} 0 & 1 & 1 & 0 \end{bmatrix}^T/\sqrt{2}, \qquad (44)$$

and its density matrix is,

$$|\beta_{01}\rangle\langle\beta_{01}| = (|01\rangle+|10\rangle)/\sqrt{2}\,(\langle 01|+\langle 10|)/\sqrt{2}$$
$$= (|01\rangle\langle 01|+|10\rangle\langle 01|+|01\rangle\langle 10|+|10\rangle\langle 10|)/2 \qquad (45)$$
$$= \frac{1}{2}\begin{bmatrix} 0 & 0 & 0 & 0 \\ 0 & 1 & 1 & 0 \\ 0 & 1 & 1 & 0 \\ 0 & 0 & 0 & 0 \end{bmatrix}.$$

Matrix of Eq. (45) exactly represents one of the cases that confirm the lay-out of Fig. 13(b). Something similar happens with two examples of non-maximally entangled states. Based on Eqs. (35, and 36) and for $|a\rangle=|0\rangle$ and $|b\rangle=|0\rangle$, the resulting state is,

$$|\gamma_{00}\rangle = u|00\rangle+v|11\rangle = (0.8536+0.3536i)|00\rangle+(0.1464-0.3536i)|11\rangle$$
$$= \begin{bmatrix} u & 0 & 0 & v \end{bmatrix}^T = \begin{bmatrix} 0.8536+0.3536i & 0 & 0 & 0.1464-0.3536i \end{bmatrix}^T, \qquad (46)$$

such that,

$$(0.8536+0.3536i)\,conj(0.8536+0.3536i)+(0.1464-0.3536i)\,conj(0.1464-0.3536i) =$$
$$(0.8536+0.3536i)(0.8536-0.3536i)+(0.1464-0.3536i)(0.1464+0.3536i) = 1, \qquad (47)$$

where $(0.8536+0.3536i) \neq (0.1464-0.3536i)$, which is why it is not a case of maximally entangled states, and $conj(\bullet)$ means *complex conjugate of* $(\bullet)$. So, the resulting density matrix is,

$$|\gamma_{00}\rangle\langle\gamma_{00}| = [(0.8536+0.3536i)|00\rangle+(0.1464-0.3536i)|11\rangle][(0.8536-0.3536i)\langle 00|+(0.1464+0.3536i)\langle 11|]$$
$$= (0.8536+0.3536i)(0.8536-0.3536i)|00\rangle\langle 00|+(0.1464-0.3536i)(0.8536-0.3536i)|11\rangle\langle 00|$$
$$+(0.8536+0.3536i)(0.1464+0.3536i)|00\rangle\langle 11|+(0.1464-0.3536i)(0.1464+0.3536i)|11\rangle\langle 11|$$
$$= \begin{bmatrix} 0.8536 & 0 & 0 & 0.3536i \\ 0 & 0 & 0 & 0 \\ 0 & 0 & 0 & 0 \\ -0.3536i & 0 & 0 & 0.1464 \end{bmatrix}.$$

$$(48)$$

As we can see in Eq. (48), the non-zero elements of this matrix occupy the same four corners as in the case $|\beta_{00}\rangle$ of Eq. (43), although with some values expressed in imaginary numbers. Whereas for $|a\rangle = |0\rangle$ and, $|b\rangle = |1\rangle$, the resulting state is,

$$|\gamma_{01}\rangle = (0.8536 + 0.3536i)|01\rangle + (0.1464 - 0.3536i)|10\rangle$$
$$= \begin{bmatrix} 0 & 0.8536 + 0.3536i & 0.1464 - 0.3536i & 0 \end{bmatrix}^T, \quad (49)$$

and its density matrix is,

$$|\gamma_{01}\rangle\langle\gamma_{01}| = \left[(0.8536 + 0.3536i)|01\rangle + (0.1464 - 0.3536i)|10\rangle\right]\left[(0.8536 - 0.3536i)\langle 01| + (0.1464 + 0.3536i)\langle 10|\right]$$
$$= (0.8536 + 0.3536i)(0.8536 - 0.3536i)|01\rangle\langle 01| + (0.1464 - 0.3536i)(0.8536 - 0.3536i)|10\rangle\langle 01|$$
$$+ (0.8536 + 0.3536i)(0.1464 + 0.3536i)|01\rangle\langle 10| + (0.1464 - 0.3536i)(0.1464 + 0.3536i)|10\rangle\langle 10|$$
$$= \begin{bmatrix} 0 & 0 & 0 & 0 \\ 0 & 0.8536 & 0.3536i & 0 \\ 0 & -0.3536i & 0.1464 & 0 \\ 0 & 0 & 0 & 0 \end{bmatrix}. \quad (50)$$

In this matrix, it can be seen that only the four central tiles are different from zero, as is the case of the Bell state $|\beta_{01}\rangle$, although with some imaginary values.

Finally, both the first-degree Fourier states $|F_{00}\rangle_2^1$ of Eq. (28), like their equivalents in Eq. (34) of third-degree, that is, $|F_{00}\rangle_2^3$, can be represented with a few cases. For example, for $|F_{00}\rangle_2^1$ where $|a\rangle = |0\rangle$ and, $|b\rangle = |0\rangle$, the corresponding Fourier state is,

$$|F_{00}\rangle_2^1 = \begin{bmatrix} 1 & 0 & (1+i)/2 & (1-i)/2 \end{bmatrix}^T / \sqrt{2}$$
$$= \left[|00\rangle + |10\rangle(1+i)/2 + |11\rangle(1-i)/2\right]/\sqrt{2} \quad (51)$$
$$= \begin{bmatrix} 0.7071 & 0 & 0.3536 + 0.3536i & 0.3536 - 0.3536i \end{bmatrix}^T,$$

and its density matrix results,

$$|F_{00}\rangle_2^1 \langle F_{00}|_2^1 = \begin{bmatrix} 0.5 & 0 & 0.25 - 0.25i & 0.25 + 0.25i \\ 0 & 0 & 0 & 0 \\ 0.25 + 0.25i & 0 & 0.25 & 0.25i \\ 0.25 - 0.25i & 0 & -0.25i & 0.25 \end{bmatrix}, \quad (52)$$

which has a layout corresponding to the case of Fig. 13(c). Something similar happens for $|a\rangle = |0\rangle$ and, $|b\rangle = |1\rangle$, where the resulting state is,

$$|F_{01}\rangle_2^1 = \begin{bmatrix} 1 & 0 & -(1+i)/2 & -(1-i)/2 \end{bmatrix}^T / \sqrt{2}$$
$$= \left[|00\rangle - |10\rangle(1+i)/2 - |11\rangle(1-i)/2\right]/\sqrt{2} \quad (53)$$
$$= \begin{bmatrix} 0.7071 & 0 & -0.3536 - 0.3536i & -0.3536 + 0.3536i \end{bmatrix}^T,$$

and its density matrix occupies the same positions as in the previous example.

$$|F_{01}\rangle_2^I \langle F_{01}|_2^I = \begin{bmatrix} 0.5 & 0 & -0.25 + 0.25i & -0.25 - 0.25i \\ 0 & 0 & 0 & 0 \\ -0.25 - 0.25i & 0 & 0.25 & 0.25i \\ -0.25 + 0.25i & 0 & -0.25i & 0.25 \end{bmatrix}. \quad (54)$$

Instead, something very different happens for $|a\rangle = |1\rangle$ and $|b\rangle = |0\rangle$, where the resulting state is,

$$\begin{aligned}|F_{10}\rangle_2^I &= \begin{bmatrix} 0 & 1 & (1-i)/2 & (1+i)/2 \end{bmatrix}^T / \sqrt{2} \\ &= \begin{bmatrix} |01\rangle + |10\rangle(1-i)/2 + |11\rangle(1+i)/2 \end{bmatrix} / \sqrt{2} \\ &= \begin{bmatrix} 0 & 0.7071 & 0.3536 - 0.3536i & 0.3536 + 0.3536i \end{bmatrix}^T, \end{aligned} \quad (55)$$

with the following density matrix,

$$|F_{10}\rangle_2^I \langle F_{10}|_2^I = \begin{bmatrix} 0 & 0 & 0 & 0 \\ 0 & 0.5 & 0.25 + 0.25i & 0.25 - 0.25i \\ 0 & 0.25 - 0.25i & 0.25 & -0.25i \\ 0 & 0.25 + 0.25i & 0.25i & 0.25 \end{bmatrix}, \quad (56)$$

which has the same non-zero elements indicated in Fig. 13(d).
(10) Finally, quantum configuration simulation platforms such as Quirk[17] present the density matrices flipped in both directions simultaneously, that is, left-to-right, and up-to-down. Aspects of this type must be taken into account when implementing any of the circuits exposed in this work. On the other hand, this same platform presents QFT modules that already have SWAP (for two qubits) and SEBQ (for more than 2 qubits) gates incorporated, both at the input and the output of the QFT modules, therefore it is not necessary to introduce them in the simulation since doing so would lead to incorrect outcomes.

*Quantum teleportation via rough entanglement*.- Quantum teleportation[8] is the first protocol created in the field of quantum communications[7], which also has a true projection on the future quantum Internet[9-14]. In the context of quantum cryptography[27], fiber optic cabling for terrestrial implementations of quantum key distribution (QKD) protocols[27] requires quantum repeaters every certain number of kilometers[28], which in turn requires a large amount of quantum memory. The problem is that the key is exposed in its passage through them. There are currently two well-defined lines of research, the first has to do with the development of quantum repeaters that do not require quantum memory, at least not that much, and the second is to replace the same quantum repeaters with some type of implementation based on quantum teleportation[8].

Taking Fig. 14 as an initial reference, this section develops both the theoretical deductions and the implementations in a simulator and an IBM Q Experience[18] 5-qubits physical machine called Lima, of the quantum teleportation protocol[8] having as a source of pairs to the three cases studied in the previous section, that is, maximally-entangled, non-maximally entangled, and rough entangled states.

Although the theoretical deductions of the three cases will be carried out using generic qubits, both the simulations and the implementations on the 5-qubit physical machine will take as an example the teleportation of computational basis states (CBS), i.e., $\{|0\rangle, |1\rangle\}$, given that being orthogonal they notably facilitate the comparison of the outcomes between the different cases of entanglement. In addition, with this type of state, it is easier to assess the internal traceability of the states (timeline) through the protocol and thus better compare the outcomes. These states with $|+\rangle = H|0\rangle$, $|-\rangle = H|1\rangle$, $|R\rangle = SH|0\rangle$, and $|L\rangle = SH|1\rangle$ are essential in quantum communications[7], in general, and QKD[27], in particular, such that if $Z$ is the phase gate[1], then $S = \sqrt{Z} = \sqrt{\begin{bmatrix} 1 & 0 \\ 0 & -1 \end{bmatrix}} = \begin{bmatrix} 1 & 0 \\ 0 & i \end{bmatrix}$.

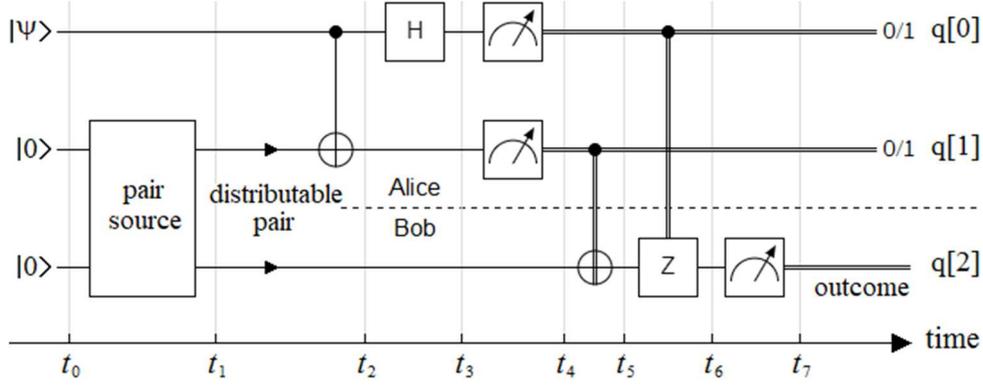

FIG. 14. Quantum teleportation protocol, where the pair-source module can emit maximally, non-maximally, and rough entangled states. The protocol starts from the side of Alice (sender) who emits the pair of particles with some of the three types of entanglement, keeping one particle and sending the other to Bob (receiver).

Based on Fig. 14, the generic state to be teleported is:

$$|\psi\rangle = \alpha|0\rangle + \beta|1\rangle, \tag{57}$$

such that, $\alpha^2 + \beta^2 = 1$. Then for $t_0$ the 3-qubits combined state is:

$$|\psi_{t_0}\rangle = |\psi 00\rangle = (\alpha|0\rangle + \beta|1\rangle)|00\rangle = \alpha|000\rangle + \beta|100\rangle. \tag{58}$$

The states of Eqs. (57, and 58) are common to the following three deductions.

*Maximally-entangled states:* In this case, the pair-source module of Fig. 14 distributes a Bell state of type $|\beta_{00}\rangle$, like that of Eq.(42). Therefore, at time $t_1$ of this figure, we will have,

$$|\psi_{t_1}\rangle = |\psi\rangle|\beta_{00}\rangle = (\alpha|0\rangle + \beta|1\rangle)\left(\frac{|00\rangle + |11\rangle}{\sqrt{2}}\right) = \frac{\alpha}{\sqrt{2}}|000\rangle + \frac{\beta}{\sqrt{2}}|100\rangle + \frac{\alpha}{\sqrt{2}}|011\rangle + \frac{\beta}{\sqrt{2}}|111\rangle \tag{59}$$

In $t_2$, a CNOT gate is applied between qubits q[0], and q[1],

$$|\psi_{t_2}\rangle = \frac{\alpha}{\sqrt{2}}|000\rangle + \frac{\beta}{\sqrt{2}}|110\rangle + \frac{\alpha}{\sqrt{2}}|011\rangle + \frac{\beta}{\sqrt{2}}|101\rangle \tag{60}$$

In $t_3$, a Hadamard ($H$) gate is applied in qubit q[0],

$$\begin{aligned}|\psi_{t_3}\rangle &= \frac{\alpha}{2}|000\rangle + \frac{\alpha}{2}|100\rangle + \frac{\beta}{2}|010\rangle - \frac{\beta}{2}|110\rangle + \frac{\alpha}{2}|011\rangle + \frac{\alpha}{2}|111\rangle + \frac{\beta}{2}|001\rangle - \frac{\beta}{2}|101\rangle \\ &= \frac{|00\rangle}{2}(\alpha|0\rangle + \beta|1\rangle) + \frac{|01\rangle}{2}(\alpha|1\rangle + \beta|0\rangle) + \frac{|10\rangle}{2}(\alpha|0\rangle - \beta|1\rangle) + \frac{|11\rangle}{2}(\alpha|1\rangle - \beta|0\rangle) \\ &= \left\{\frac{|00\rangle}{2}X^0Z^0|\psi\rangle\right\} + \left\{\frac{|01\rangle}{2}X^1Z^0|\psi\rangle\right\} + \left\{\frac{|10\rangle}{2}X^0Z^1|\psi\rangle\right\} + \left\{\frac{|11\rangle}{2}X^1Z^1|\psi\rangle\right\}. \\ &\quad\ 25\%\qquad\qquad\qquad 25\%\qquad\qquad\qquad 25\%\qquad\qquad\qquad 25\%\end{aligned} \tag{61}$$

All the terms of the last row of Eq.(61) have the same probability, that is, 25%, since the four bases $\{|00\rangle,|01\rangle,|10\rangle,|11\rangle\}$ are equiprobable at the output of the Bell-State-Measurement (BSM) module[1], which is between the qubits q[0] and q[1] and is composed from the gates *CNOT*, *H* and the two quantum measurement blocks. This happens at $t_4$, where the exponents of the *X* and *Z* matrices are the classical bits of disambiguation needed to reconstruct the teleported state on Bob's (receiver) side. As a direct consequence of this, Bob must apply a gate *X* at $t_5$, and a gate *Z* at $t_6$, if the respective disambiguation bits have a value equal to 1. Thus, the rebuilt outcome is obtained in $t_7$ after the quantum measurement in q[2].

Now, if $|\psi\rangle = \begin{bmatrix}\alpha\\ \beta\end{bmatrix} = \begin{bmatrix}1\\ 0\end{bmatrix} = |0\rangle$, and replacing $\alpha$ and $\beta$ in Eq.(61) at time $t_3$, we will have,

$$\begin{aligned}|\psi_{t_3}\rangle &= \frac{1}{2}|000\rangle + \frac{1}{2}|100\rangle + \frac{1}{2}|011\rangle + \frac{1}{2}|111\rangle \\ &= \frac{|00\rangle}{2}|0\rangle + \frac{|01\rangle}{2}|1\rangle + \frac{|10\rangle}{2}|0\rangle + \frac{|11\rangle}{2}|1\rangle \\ &= \left\{\begin{array}{c}\frac{|00\rangle}{2}X^0Z^0|0\rangle\\ 25\%\end{array}\right\} + \left\{\begin{array}{c}\frac{|01\rangle}{2}X^1Z^0|0\rangle\\ 25\%\end{array}\right\} + \left\{\begin{array}{c}\frac{|10\rangle}{2}X^0Z^1|0\rangle\\ 25\%\end{array}\right\} + \left\{\begin{array}{c}\frac{|11\rangle}{2}X^1Z^1|0\rangle\\ 25\%\end{array}\right\}.\end{aligned} \quad (62)$$

From Eq.(62), for $|0\rangle$, the sum of the probabilities is 25% + 25% + 25% + 25% = 100%, while for $|1\rangle$, the sum of the probabilities is 0%.

Instead, if $|\psi\rangle = \begin{bmatrix}\alpha\\ \beta\end{bmatrix} = \begin{bmatrix}0\\ 1\end{bmatrix} = |1\rangle$, replacing $\alpha$ and $\beta$ in Eq.(61) at time $t_3$, and taking into account that $Z|0\rangle = |0\rangle$, but $Z|1\rangle = -|1\rangle$, it turns out that,

$$\begin{aligned}|\psi_{t_3}\rangle &= \frac{1}{2}|010\rangle - \frac{1}{2}|110\rangle + \frac{1}{2}|001\rangle - \frac{1}{2}|101\rangle \\ &= \frac{|00\rangle}{2}|1\rangle + \frac{|01\rangle}{2}|0\rangle + \frac{|10\rangle}{2}(-|1\rangle) + \frac{|11\rangle}{2}(-|0\rangle) \\ &= \left\{\begin{array}{c}\frac{|00\rangle}{2}X^0Z^0|1\rangle\\ 25\%\end{array}\right\} + \left\{\begin{array}{c}\frac{|01\rangle}{2}X^1Z^0|1\rangle\\ 25\%\end{array}\right\} + \left\{\begin{array}{c}\frac{|10\rangle}{2}X^0Z^1|1\rangle\\ 25\%\end{array}\right\} + \left\{\begin{array}{c}\frac{|11\rangle}{2}X^1Z^1|1\rangle\\ 25\%\end{array}\right\}\end{aligned} \quad (63)$$

From Eq.(63), for $|0\rangle$, the sum of the probabilities is 0%, while for $|1\rangle$, the sum of the probabilities is 25% + 25% + 25% + 25% = 100%.

*Non-maximally-entangled states:* Given the state $|\gamma_{00}\rangle$ of Eq. (46) but in a general way, i.e., with generic *u* and *v*, and considering that for instant $t_0$ this case is the same as the previous one, at $t_1$, it results,

$$|\psi_{t_1}\rangle = |\psi\rangle|\gamma_{00}\rangle = (\alpha|0\rangle + \beta|1\rangle)(u|00\rangle + v|11\rangle) = \alpha u|000\rangle + \beta u|100\rangle + \alpha v|011\rangle + \beta v|111\rangle, \quad (64)$$

while at time $t_2$ a *CNOT* gate is applied between the qubits q[0] and q[1]:

$$|\psi_{t_2}\rangle = \alpha u|000\rangle + \beta u|110\rangle + \alpha v|011\rangle + \beta v|101\rangle. \quad (65)$$

At instant $t_3$, a Hadamard gate ($H$) is applied in qubit q[0],

$$|\psi_{t_3}\rangle = \frac{\alpha u|000\rangle + \alpha u|100\rangle + \beta u|010\rangle - \beta u|110\rangle + \alpha v|011\rangle + \alpha v|111\rangle + \beta v|001\rangle - \beta v|101\rangle}{\sqrt{2}}$$

$$= \frac{|00\rangle}{\sqrt{2}}(\alpha u|0\rangle + \beta v|1\rangle) + \frac{|01\rangle}{\sqrt{2}}(\alpha v|1\rangle + \beta u|0\rangle) + \frac{|10\rangle}{\sqrt{2}}(\alpha u|0\rangle - \beta v|1\rangle) + \frac{|11\rangle}{\sqrt{2}}(\alpha v|1\rangle - \beta u|0\rangle) \quad (66)$$

$$= \left\{\begin{array}{c}\frac{|00\rangle}{\sqrt{2}}X^0 Z^0|\overline{\psi}\rangle \\ 25\%\end{array}\right\} + \left\{\begin{array}{c}\frac{|01\rangle}{\sqrt{2}}X^1 Z^0|\overline{\psi}\rangle \\ 25\%\end{array}\right\} + \left\{\begin{array}{c}\frac{|10\rangle}{\sqrt{2}}X^0 Z^1|\overline{\psi}\rangle \\ 25\%\end{array}\right\} + \left\{\begin{array}{c}\frac{|11\rangle}{\sqrt{2}}X^1 Z^1|\overline{\psi}\rangle \\ 25\%\end{array}\right\},$$

where $|\overline{\psi}\rangle = \alpha u|0\rangle + \beta v|1\rangle$. The last line of Eq.(66) shows that the four terms are equiprobable for the state $|\overline{\psi}\rangle$, with a 25% probability for each of the four bases. However, since $u$ and $v$ are generally different, as seen in Eqs. (35, and 46), this causes an imbalance in the probability distribution concerning the four bases, due to the crossing between the coefficients $\alpha$, $\beta$, $u$, and $v$. In practice, this type of entanglement does not facilitate the teleportation of any type of state $|\psi\rangle$, this being another reason why we resort to the CBS teleportation as an example, so in this way, it is possible to compare the outcomes for the three cases of entanglement. As far as times $t_4$, $t_5$, $t_6$, and $t_7$ are concerned, a similar description of the case of maximally-entangled states takes place.

As for the previous case, if $|\psi\rangle = \begin{bmatrix}\alpha\\\beta\end{bmatrix} = \begin{bmatrix}1\\0\end{bmatrix} = |0\rangle$, we can now replace $\alpha$ and $\beta$ in Eq.(66) at time $t_3$, however, we must take into account Eqs. (35, 36, and 46), in which case we consider that $\sqrt[4]{X} = HTH$, where $T = \sqrt[4]{Z} = \begin{bmatrix}1 & 0\\0 & e^{i\frac{\pi}{4}}\end{bmatrix}$. This replacement will make it possible to implement this particular case of entanglement on the selected IBM Q[18] machine, since the aforementioned platform does not have the gate $\sqrt[4]{X}$ in its arsenal. Therefore, we will have,

$$|\psi_{t_3}\rangle = \frac{u|000\rangle + u|100\rangle + v|011\rangle + v|111\rangle}{\sqrt{2}}$$

$$= \frac{|00\rangle}{\sqrt{2}}u|0\rangle + \frac{|01\rangle}{\sqrt{2}}v|1\rangle + \frac{|10\rangle}{\sqrt{2}}u|0\rangle + \frac{|11\rangle}{\sqrt{2}}v|1\rangle \quad (67)$$

$$= \left\{\begin{array}{c}\frac{|00\rangle}{\sqrt{2}}X^0 Z^0 u|0\rangle \\ 42.68\%\end{array}\right\} + \left\{\begin{array}{c}\frac{|01\rangle}{\sqrt{2}}X^1 Z^0 v|0\rangle \\ 7.32\%\end{array}\right\} + \left\{\begin{array}{c}\frac{|10\rangle}{\sqrt{2}}X^0 Z^1 u|0\rangle \\ 42.68\%\end{array}\right\} + \left\{\begin{array}{c}\frac{|11\rangle}{\sqrt{2}}X^1 Z^1 v|0\rangle \\ 7.32\%\end{array}\right\}$$

The very particular distribution of probabilities of the last line of Eq.(67) arises from:
$u\ conj(u)/2 = (0.8536 + 0.3536i)\ conj(0.8536 + 0.3536i)/2 = 0.4268 \rightarrow 42.68\%$, and
$v\ conj(v)/2 = (0.1464 - 0.3536i)\ conj(0.1464 - 0.3536i)/2 = 0.732 \rightarrow 7.32\%$.

From Eq.(67), for $|0\rangle$, the sum of the probabilities is $42.68\% + 7.32\% + 42.68\% + 7.32\% = 100\%$, while for $|1\rangle$, the sum of the probabilities results $0\%$.

Instead, if $|\psi\rangle = \begin{bmatrix}\alpha\\\beta\end{bmatrix} = \begin{bmatrix}0\\1\end{bmatrix} = |1\rangle$, replacing $\alpha$ and $\beta$ in Eq.(66) at time $t_3$, and bearing in mind again that $Z|0\rangle = |0\rangle$, but $Z|1\rangle = -|1\rangle$, it turns out that,

$$|\psi_{t_3}\rangle = \frac{u|010\rangle - u|110\rangle + v|001\rangle - v|101\rangle}{\sqrt{2}}$$

$$= \frac{|00\rangle}{\sqrt{2}} v|1\rangle + \frac{|01\rangle}{\sqrt{2}} u|0\rangle + \frac{|10\rangle}{\sqrt{2}} (-v|1\rangle) + \frac{|11\rangle}{\sqrt{2}} (-u|0\rangle) \qquad (68)$$

$$= \left\{ \underbrace{\frac{|00\rangle}{\sqrt{2}} X^0 Z^0 v|1\rangle}_{7.32\%} \right\} + \left\{ \underbrace{\frac{|01\rangle}{\sqrt{2}} X^1 Z^0 u|1\rangle}_{42.68\%} \right\} + \left\{ \underbrace{\frac{|10\rangle}{\sqrt{2}} X^0 Z^1 v|1\rangle}_{7.32\%} \right\} + \left\{ \underbrace{\frac{|11\rangle}{\sqrt{2}} X^1 Z^1 u|1\rangle}_{42.68\%} \right\}.$$

From Eq.(68), for $|0\rangle$, the sum of the probabilities is 0%, while for $|1\rangle$, the sum of the probabilities is 7.32% + 42.68% + 7.32% + 42.68% = 100%, that is, an identical result to the case of maximally-entangled states, although for a very particular state to be teleported as is the case of $|0\rangle$.

*Rough-entangled states:* From Eq.(51) we obtain $|F_{00}\rangle_2^1$, then at time $t_1$, we have,

$$|\psi_{t_1}\rangle = |\psi\rangle |F_{00}\rangle_2^1 = (\alpha|0\rangle + \beta|1\rangle) \left( \frac{|00\rangle + |10\rangle(1+i)/2 + |11\rangle(1-i)/2}{\sqrt{2}} \right) \qquad (69)$$

$$= \frac{\alpha}{\sqrt{2}} |000\rangle + \frac{\beta}{\sqrt{2}} |100\rangle + \frac{\alpha}{\sqrt{2}} \frac{(1+i)}{2} |010\rangle + \frac{\beta}{\sqrt{2}} \frac{(1+i)}{2} |110\rangle + \frac{\alpha}{\sqrt{2}} \frac{(1-i)}{2} |011\rangle + \frac{\beta}{\sqrt{2}} \frac{(1-i)}{2} |111\rangle,$$

while at time $t_2$ a *CNOT* gate is applied between qubits q[0] and q[1],

$$|\psi_{t_2}\rangle = \frac{\alpha}{\sqrt{2}} |000\rangle + \frac{\beta}{\sqrt{2}} |110\rangle + \frac{\alpha}{\sqrt{2}} \frac{(1+i)}{2} |010\rangle + \frac{\beta}{\sqrt{2}} \frac{(1+i)}{2} |100\rangle + \frac{\alpha}{\sqrt{2}} \frac{(1-i)}{2} |011\rangle + \frac{\beta}{\sqrt{2}} \frac{(1-i)}{2} |101\rangle. \qquad (70)$$

At time $t_3$, a Hadamard (*H*) gate is applied in qubit q[0],

$$|\psi_{t_3}\rangle = \frac{\alpha}{2} |000\rangle + \frac{\alpha}{2} |100\rangle + \frac{\beta}{2} |010\rangle - \frac{\beta}{2} |110\rangle + \frac{\alpha}{2} \frac{(1+i)}{2} |010\rangle + \frac{\alpha}{2} \frac{(1+i)}{2} |110\rangle + \frac{\beta}{2} \frac{(1+i)}{2} |000\rangle$$

$$- \frac{\beta}{2} \frac{(1+i)}{2} |100\rangle + \frac{\alpha}{2} \frac{(1-i)}{2} |011\rangle + \frac{\alpha}{2} \frac{(1-i)}{2} |111\rangle + \frac{\beta}{2} \frac{(1-i)}{2} |001\rangle - \frac{\beta}{2} \frac{(1-i)}{2} |101\rangle$$

$$= |00\rangle \left\{ \underbrace{\left\{ \frac{|0\rangle}{2} \left[ \alpha + \beta \frac{(1+i)}{2} \right] \right\}}_{12.5\%} + \underbrace{\left\{ \frac{|1\rangle}{2} \left[ \beta \frac{(1-i)}{2} \right] \right\}}_{12.5\%} \right\} + |10\rangle \left\{ \underbrace{\left\{ \frac{|0\rangle}{2} \left[ \alpha - \beta \frac{(1+i)}{2} \right] \right\}}_{12.5\%} + \underbrace{\left\{ \frac{|1\rangle}{2} \left[ -\beta \frac{(1-i)}{2} \right] \right\}}_{12.5\%} \right\} \qquad (71)$$

$$+ |01\rangle \left\{ \underbrace{\left\{ \frac{|0\rangle}{2} \left[ \alpha \frac{(1+i)}{2} + \beta \right] \right\}}_{12.5\%} + \underbrace{\left\{ \frac{|1\rangle}{2} \left[ \alpha \frac{(1-i)}{2} \right] \right\}}_{12.5\%} \right\} + |11\rangle \left\{ \underbrace{\left\{ \frac{|0\rangle}{2} \left[ \alpha \frac{(1+i)}{2} - \beta \right] \right\}}_{12.5\%} + \underbrace{\left\{ \frac{|1\rangle}{2} \left[ \alpha \frac{(1-i)}{2} \right] \right\}}_{12.5\%} \right\}.$$

The eight final terms of Eq.(71) show us a distribution of probabilities quite fragmented due to the intervention of complex coefficients because of a single intervention of the QFT. However, for each of the four bases the associated final probability is 25%. Similar considerations to the previous cases take place at times $t_4$, $t_5$, $t_6$, and $t_7$. Moreover, very similar results would be obtained using $|F_{00}\rangle_2^3$ instead of $|F_{00}\rangle_2^1$.

As for the two previous cases, if $|\psi\rangle = \begin{bmatrix} \alpha \\ \beta \end{bmatrix} = \begin{bmatrix} 1 \\ 0 \end{bmatrix} = |0\rangle$, we can replace $\alpha$ and $\beta$ in Eq.(71) at time $t_3$, where we get,

$$|\psi_{t_3}\rangle = |00\rangle\frac{|0\rangle}{2} + |10\rangle\frac{|0\rangle}{2} + |01\rangle\left[\frac{|0\rangle}{2}\frac{(1+i)}{2} + \frac{|1\rangle}{2}\frac{(1-i)}{2}\right] + |11\rangle\left[\frac{|0\rangle}{2}\frac{(1+i)}{2} + \frac{|1\rangle}{2}\frac{(1-i)}{2}\right]$$

$$= |00\rangle X^0 Z^0 \frac{|0\rangle}{2} + |01\rangle X^1 Z^0 \left[\frac{|1\rangle}{2}\frac{(1+i)}{2} + \frac{|0\rangle}{2}\frac{(1-i)}{2}\right]$$

$$+ |10\rangle X^0 Z^1 \frac{|0\rangle}{2} + |11\rangle X^1 Z^1 \left[\frac{-|1\rangle}{2}\frac{(1+i)}{2} + \frac{|0\rangle}{2}\frac{(1-i)}{2}\right] \quad (72)$$

$$= |00\rangle\left\{\begin{matrix}X^0 Z^0|0\rangle/2\\25\%\end{matrix}\right\} + |01\rangle\left[\left\{\begin{matrix}X^1 Z^0|1\rangle(1+i)/4\\12.5\%\end{matrix}\right\} + \left\{\begin{matrix}X^1 Z^0|0\rangle(1-i)/4\\12.5\%\end{matrix}\right\}\right]$$

$$+ |10\rangle\left\{\begin{matrix}X^0 Z^1|0\rangle/2\\25\%\end{matrix}\right\} + |11\rangle\left[\left\{\begin{matrix}-X^1 Z^1|1\rangle(1+i)/4\\12.5\%\end{matrix}\right\} + \left\{\begin{matrix}X^1 Z^1|0\rangle(1-i)/4\\12.5\%\end{matrix}\right\}\right].$$

The probabilities of the last terms of Eq.(72) arise from:
$\frac{1}{2}\frac{1}{2} = \frac{1}{4} \to 25\%$, while $\frac{1}{4}\frac{2}{\sqrt{2}}\frac{1}{4}\frac{2}{\sqrt{2}} = \frac{4}{32} = \frac{1}{8} \to 12.5\%$.

Then, from Eq.(72), for $|0\rangle$, the sum of the probabilities is 25% + 12.5% + 25% + 12.5% = 75%, while for $|1\rangle$, the sum of the probabilities is 12.5% + 12.5% = 25%.

Instead, if $|\psi\rangle = \begin{bmatrix}\alpha\\\beta\end{bmatrix} = \begin{bmatrix}0\\1\end{bmatrix} = |1\rangle$, replacing $\alpha$ and $\beta$ in Eq.(71) at time $t_3$, and taking into account again that $Z|0\rangle = |0\rangle$, but $Z|1\rangle = -|1\rangle$, it turns out that,

$$|\psi_{t_3}\rangle = |00\rangle\left[\frac{|0\rangle}{2}\frac{(1+i)}{2} + \frac{|1\rangle}{2}\frac{(1-i)}{2}\right] + |10\rangle\left[-\frac{|0\rangle}{2}\frac{(1+i)}{2} - \frac{|1\rangle}{2}\frac{(1-i)}{2}\right] + |01\rangle\frac{|0\rangle}{2} - |11\rangle\frac{|0\rangle}{2}$$

$$= |00\rangle X^0 Z^0\left[\frac{|0\rangle}{2}\frac{(1+i)}{2} + \frac{|1\rangle}{2}\frac{(1-i)}{2}\right] + |01\rangle X^1 Z^0 \frac{|1\rangle}{2}$$

$$+ |10\rangle X^0 Z^1\left[-\frac{|0\rangle}{2}\frac{(1+i)}{2} + \frac{|1\rangle}{2}\frac{(1-i)}{2}\right] + |11\rangle X^1 Z^1 \frac{|1\rangle}{2} \quad (73)$$

$$= |00\rangle\left[\left\{\begin{matrix}X^0 Z^0|0\rangle(1+i)/4\\12.5\%\end{matrix}\right\} + \left\{\begin{matrix}X^0 Z^0|1\rangle(1-i)/4\\12.5\%\end{matrix}\right\}\right] + |01\rangle\left\{\begin{matrix}X^1 Z^0|1\rangle/2\\25\%\end{matrix}\right\} +$$

$$+ |10\rangle\left[\left\{\begin{matrix}-X^0 Z^1|0\rangle(1+i)/4\\12.5\%\end{matrix}\right\} + \left\{\begin{matrix}X^0 Z^1|1\rangle(1-i)/4\\12.5\%\end{matrix}\right\}\right] + |11\rangle\left\{\begin{matrix}X^1 Z^1|1\rangle/2\\25\%\end{matrix}\right\}.$$

Therefore, from Eq.(73), for $|0\rangle$, the sum of the probabilities is 12.5% + 12.5% = 25%, while for $|1\rangle$, the sum of the probabilities is 25% + 12.5% + 25% + 12.5% = 75%.

From these two examples, that is, when the state to be teleported is a CBS, we see that the outcomes have a probability of 75% for the teleported state, and 25% for its counterpart. As will be seen below in the implementations that will take place on both platforms of IBM Q[18], i.e., simulator and Lima 5-qubit physical machine, this will not constitute a problem at all in terms of the discrimination of both states after the process of obtaining the outcome from the quantum measurement, given that the difference between both probabilities is twice the smallest of them.

*Results in the IBM $Q^{18}$ simulator:* Teleportations of both CBS are carried out for all three types of entanglement on this platform. However, before starting, a point of fundamental importance must be clarified to understand the implementations and conveniently contrast these implementations with the theoretical deductions of the previous sections. This point consists of the following: in the abscissa axis of the histograms obtained both in the simulator and in the IBM $Q^{18}$ physical machines, the qubits are always shown in the following order: q[2]q[1]q[0], where q[2] is found at the bottom of Fig. 14, and q[0] at the top of it, while in the theoretical derivations of Eqs. (61-63, 66-68, and 71-73), the order is exactly the opposite, i.e., q[0]q[1]q[2]. However, for both criteria, q[2] is the qubit under analysis, while q[1] and q[0] constitute the base present in the Bell State Measurement (BSM) module. With this important point clear, the simulations begin in Fig. 15, where figures (a) for $|0\rangle$, and (b) for $|1\rangle$ show the histograms resulting from working with maximally-entangled states, while figures (c) for $|0\rangle$, and (d) for $|1\rangle$ contain the histograms relative to the non-maximally entangled states. Finally, the figures (e) for $|0\rangle$, and (f) for $|1\rangle$ represent the histograms when teleportation is carried out using rough entangled states. Both in figures (a) and (c) the percentages are 100% for $|0\rangle$ and 0% for $|1\rangle$, while in figures (b) and (d) the opposite occurs, as was deduced theoretically. Finally, in the figures (e, and f) corresponding to $|F_{00}\rangle_2^1$, the four measurement bases are involved as predicted in the theoretical deduction with 75% for $|0\rangle$ and 25% for $|1\rangle$ for the first case, and the opposite for the second one.

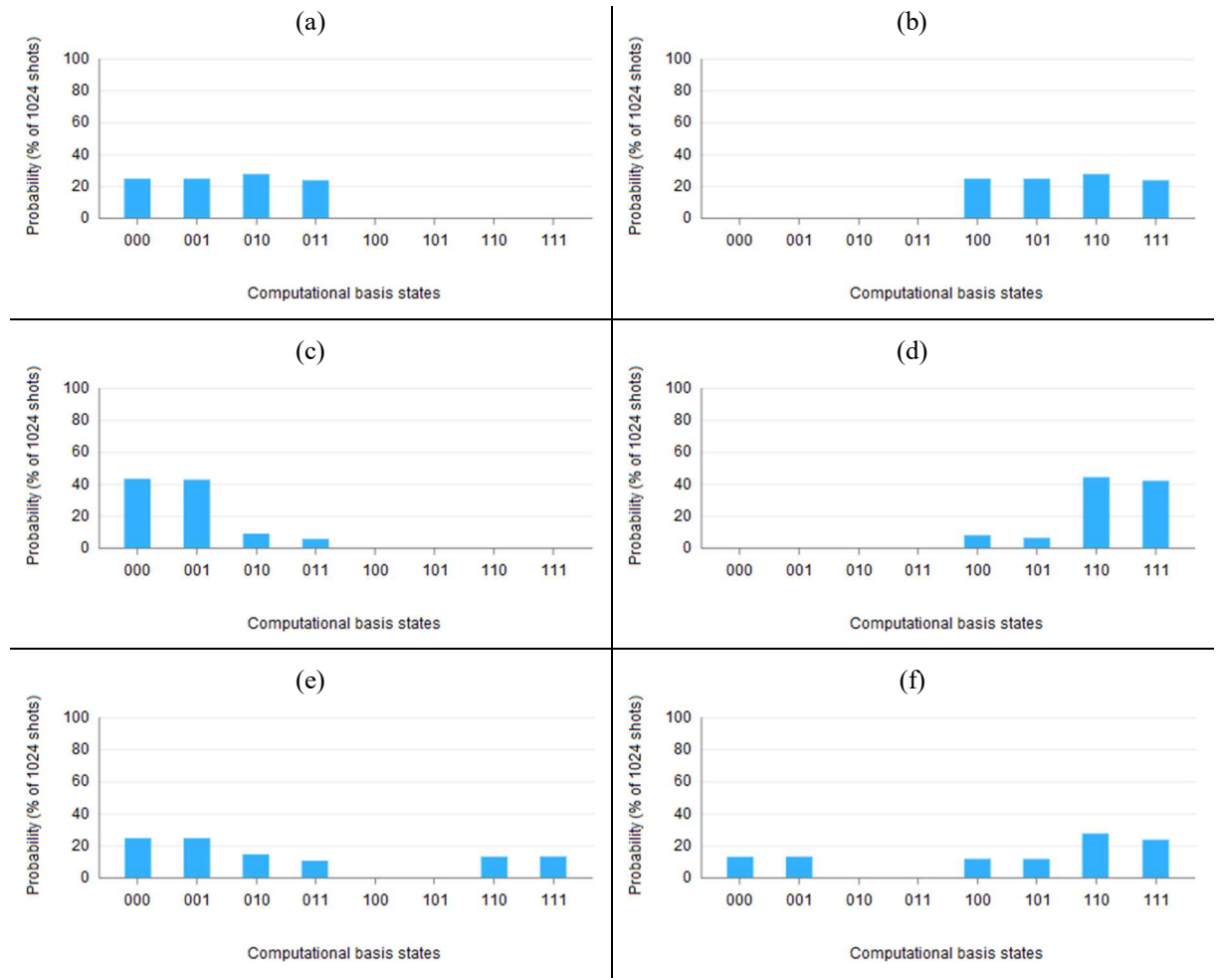

FIG. 15. Histograms of the teleportations for the three types of entanglement. The top row corresponds to maximally entangled states, the middle row to non-maximally entangled states, and the bottom row corresponds to rough-entangled states. The left column results from teleporting $|\psi\rangle = |0\rangle$, while the right column results from teleporting $|\psi\rangle = |1\rangle$.

*Results in the IBM $Q^{18}$ Lima 5-qubits processor:* For these implementations, we will resort to the simplified version of the quantum teleportation protocol of Fig. 16, i.e., the one without quantum measurement modules in the qubits q[0] and q[1]. This is because the physical machines of IBM $Q^{18}$, as in the case of Lima, do not allow quantum measurement modules in intermediate instances of the quantum circuit, so we resort to the simplified version of the protocol shown in Fig. 16.

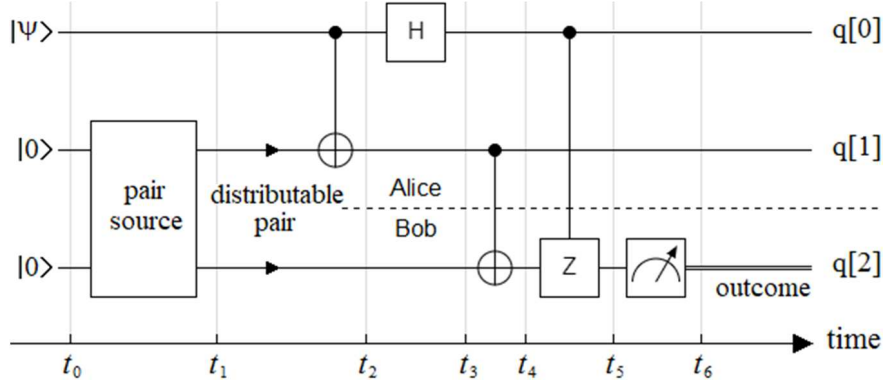

FIG. 16. A simplified version of the quantum teleportation protocol without quantum measurement modules in qubits q[0], and q[1], allows the implementation of this protocol in any physical machine of IBM $Q^{18}$, for example in one of the 5-qubits like Lima.

In Fig. 17, figures (a) for $|0\rangle$, and (b) for $|1\rangle$ show the histograms resulting from working with maximally-entangled states, while figures (c) for $|0\rangle$, and (d) for $|1\rangle$ contain the histograms relative to the non-maximally entangled states. The decoherence present in every physical machine, of which Lima is no exception, makes non-zero probabilities appear where it does not correspond, both when we work with maximally entangled states and in the case of non-maximally entangled states. Both in figures (a) and (c) the percentages are not 100% for $|0\rangle$, and are not 0% for $|1\rangle$, while in figures (b) and (d) the opposite occurs. This contrasts both with what was deduced theoretically and with the results obtained in the simulator. On the other hand, figures (e) for $|0\rangle$, and (f) for $|1\rangle$ represent the histograms when teleportation is carried out using rough entangled states. In these figures, corresponding to $\left|F_{00}\right\rangle_{2}^{1}$, the four measurement bases are involved as predicted in the theoretical deduction, and although the results are not exactly 75% for $|0\rangle$ and 25% for $|1\rangle$ for the first case, and the opposite for the second. However, as will be seen in the next section, the case of rough entangled states has the smallest absolute value of the three types of entanglement.

*Analysis of the results:* Table 10 shows that the theoretical predictions were carried out due to the deduction of Eqs. (62, 63, 67, 68, 72, and 73), for the three types of entanglement when the state to be teleported is a CBS, fit satisfactorily with the experimental results obtained both in the simulator and in the 5-qubit Lima processor of IBM $Q^{18}$. Table 11 represents the correspondence between the qubit to be teleported, the obtained outcome, and the post-processing required by the rough entanglement case to be useful in the context of the future quantum Internet[9-14]. This post-processing does not represent any reduction in the performance of the teleportation protocol, since its application does not imply the use of any type of special technology to achieve it, since it must be carried out after the quantum measurement of the qubit q[2], both in the configuration of Fig. 14 and 16, that is, once the wave function has collapsed, or what is the same, in the classical world.

Notwithstanding what has been said, the absolute outcome error, i.e., the difference between the theoretical values and those obtained in the 5-qubit Lima processor of IBM $Q^{18}$, will always be lower in the case of working with rough entangled states. For example, if the qubit to be teleported is $|\psi\rangle = |0\rangle$, for all three entanglement types, that absolute outcome error is:

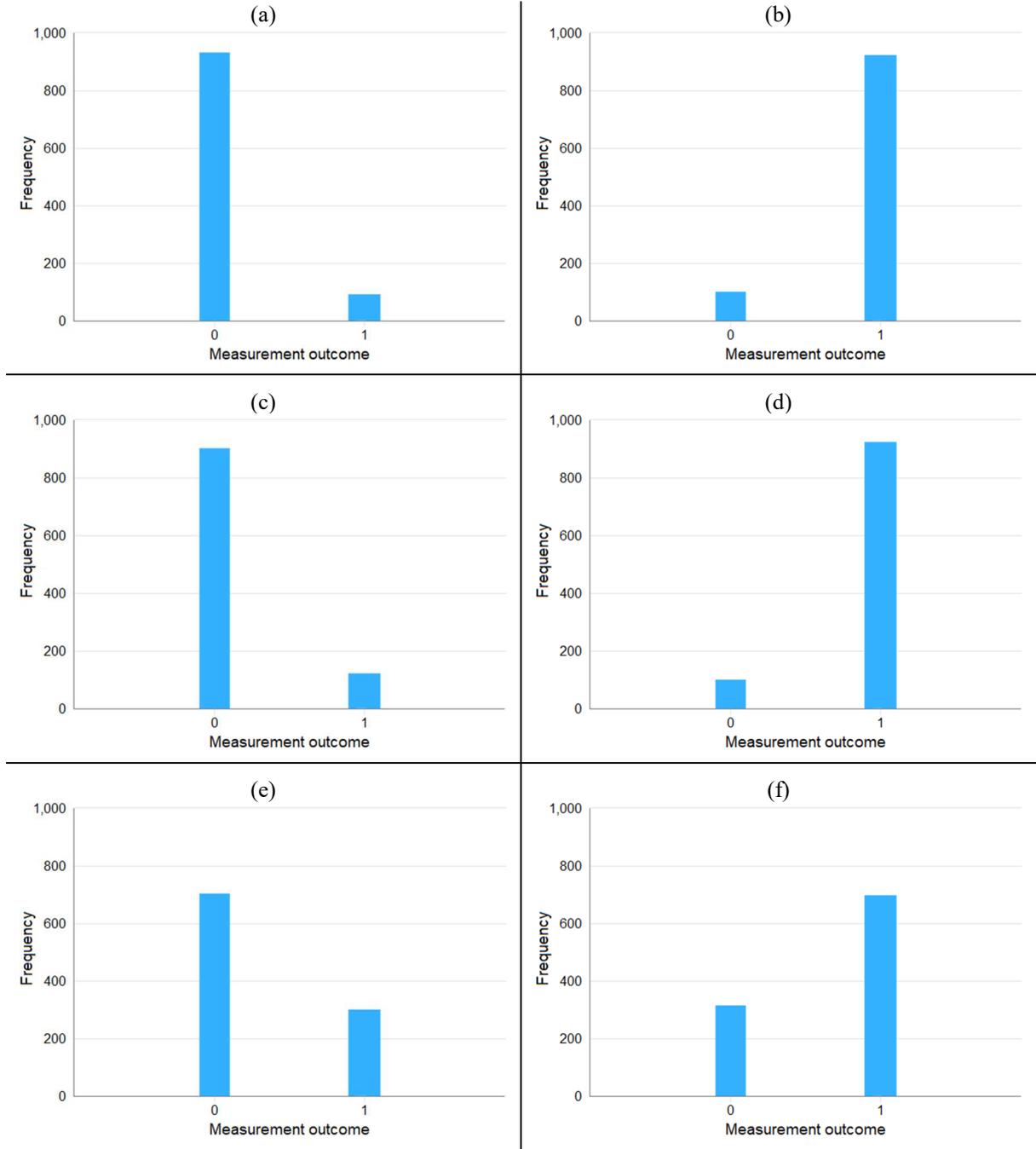

FIG. 17. Histograms of the teleportations for the three types of entanglement. The top row corresponds to maximally entangled states, the middle row to non-maximally entangled states, and the bottom row corresponds to rough-entangled states. The left column results from teleporting $|\psi\rangle = |0\rangle$, while the right column results from teleporting $|\psi\rangle = |1\rangle$.

*Maximally-entangled states:*
$$\Delta_0 = |P_{0,theoretical} - P_{0,Lima}| = 1 - 0.9101 = 0.0899, \tag{74}$$

*Non-maximally-entangled states:*
$$\Delta_0 = |P_{0,theoretical} - P_{0,Lima}| = 1 - 0.8808 = 0.1192, \tag{75}$$

*Rough-entangled states:*
$$\Delta_0 = |P_{0,theoretical} - P_{0,Lima}| = 0.75 - 0.6972 = 0.0528. \tag{76}$$

| States → | Maximally-entangled | | Non-maximally entangled | | Rough-entangled | |
|---|---|---|---|---|---|---|
| $|\psi\rangle \rightarrow$ | $|0\rangle$ | $|1\rangle$ | $|0\rangle$ | $|1\rangle$ | $|0\rangle$ | $|1\rangle$ |
| Theoretical Prob. of $|0\rangle$ | 1 | 0 | 1 | 0 | 0.75 | 0.25 |
| Theoretical Prob. of $|1\rangle$ | 0 | 1 | 0 | 1 | 0.25 | 0.75 |
| Simulator Prob. of $|0\rangle$ | 1 | 0 | 1 | 0 | 0.7392 | 0.2588 |
| Simulator Prob. of $|1\rangle$ | 0 | 1 | 0 | 1 | 0.2608 | 0.7412 |
| Lima Prob. of $|0\rangle$ | 0.9101 | 0.0986 | 0.8808 | 0.0976 | 0.6972 | 0.3184 |
| Lima Prob. of $|1\rangle$ | 0.0899 | 0.9014 | 0.1192 | 0.9024 | 0.3028 | 0.6816 |

Table 10. Outcome comparison for three types of entanglement among the theoretical deductions, the simulator, and the 5-qubit Lima processor.

| qubit to be teleported | outcome | after post-processing |
|---|---|---|
| $|0\rangle$ | $\{75\%|0\rangle, 25\%|1\rangle\} \rightarrow$ | 0 |
| $|1\rangle$ | $\{25\%|0\rangle, 75\%|1\rangle\} \rightarrow$ | 1 |

Table 11. Outcome post-processing in teleportation with rough entanglement.

From Eqs. (74-76), it follows that the case of non-maximally-entangled-states has the highest absolute-outcome-error, while rough entangled states have the lowest one. As we have seen, the compilation of probabilities, the terms of Eq. (71) to reconstruct the outcome, is more rustic for rough entangled states, however, the difference in probabilities is greater than the smallest probability, i.e., 75% - 25% = 50% > 25%. For this reason, although both possible outcomes are not orthogonal to each other (as in the other two cases), there is enough discrimination between them to be distinguished.

When choosing to teleport computational basis states (CBS) $\{|0\rangle, |1\rangle\}$, which constitute the mutually orthogonal pair par excellence, it is possible to evaluate decoherence introduced by the platform that hosts the experiment better than with any other qubits. Moreover, the use of CBS as qubits to be teleported makes it easier to compare the three entanglement cases than any other pair of qubits.

All implementations of this study are available in both Quirk[17] and IBM Q[18] in the *Data Availability Statement* section.

Finally, similar results to those obtained in this section would be obtained using $|F_{00}\rangle_2^3$ instead of $|F_{00}\rangle_2^1$, the reason why the same implementations with $|F_{00}\rangle_2^3$ are not repeated.

***Other applications.-*** In this section, five of the most conspicuous cases of quantum-Fourier-gates (QFG) application have been selected to be developed, which are:
- Quantum stretching,
- entanglement levels,
- entanglement parallelization,
- quantum secret sharing[29] (QSS) for quantum cryptography[27], and
- quantum repeaters[16] for QKD[28], and the future quantum Internet[9-14].

These techniques will have a great projection on quantum communication and cryptography, and their development here constitutes only a small part of the universe of QFG applications. Moreover, without loss of generality, for the first three techniques it was decided to explain them for configurations of no more than four qubits, while for the last two, we will resort to the same criterion, but for three, and two qubits, respectively. This is why a greater number of qubits would cause a great increase in the number, as well as in the size of the associated figures necessary to explain them, which would inappropriately extend the dimension of this work.

*Quantum stretching:* It specifically consists of a detailed analysis of the dimensional transition between entanglement configurations for a consecutive number of qubits, where the mentioned transition is regulated by the subscript of the $F_k^2$ gate that accompanies the Hadamard gate (H). That is, by changing the subscript of the $F_k^2$ gate, the degree of stretching is changed, $k$ being the stretching index. Then, Eqs. (77-80) show the mentioned transition for incremental values of the subscript $k$, which can be seen in detail in the second lines of each equation, where the stretching is manifested as the inclusion of numerous zeros in the middle of the ones that are found at the ends. In other words, the coefficients involved are the same, it is simply stretching of the respective vectors, which is modeled by a simple change in a parameter of the $F_k^2$ gate, which unites the entire family of entangled particles.

$$|F_0\rangle_1^2 = |+\rangle = F_1^2 H|0\rangle = (|0\rangle + |1\rangle)/\sqrt{2}$$
$$= \begin{bmatrix} 1 & 1 \end{bmatrix}^T / \sqrt{2} \tag{77}$$

$$|F_{00}\rangle_2^2 = |\beta_{00}\rangle = F_2^2 \left(H \otimes I_{2^1 \times 2^1}\right)|00\rangle = (|00\rangle + |11\rangle)/\sqrt{2}$$
$$= \begin{bmatrix} 1 & 0 & 0 & 1 \end{bmatrix}^T / \sqrt{2} \tag{78}$$

$$|F_{000}\rangle_3^2 = |GHZ_3\rangle = F_3^2 \left(H \otimes I_{2^2 \times 2^2}\right)|000\rangle = (|000\rangle + |111\rangle)/\sqrt{2}$$
$$= \begin{bmatrix} 1 & 0 & 0 & 0 & 0 & 0 & 0 & 1 \end{bmatrix}^T / \sqrt{2} \tag{79}$$

$$|F_{0000}\rangle_4^2 = |GHZ_4\rangle = F_4^2 \left(H \otimes I_{2^3 \times 2^3}\right)|0000\rangle = (|0000\rangle + |1111\rangle)/\sqrt{2}$$
$$= \begin{bmatrix} 1 & 0 & 0 & 0 & 0 & 0 & 0 & 0 & 0 & 0 & 0 & 0 & 0 & 0 & 0 & 1 \end{bmatrix}^T / \sqrt{2} \tag{80}$$

In Fig. 18, the four cases of Eqs. (77-80) are represented using a correlative transition that goes from $|+\rangle$ to $|GHZ_4\rangle$, passing through $|GHZ_2\rangle$ and $|GHZ_3\rangle$.

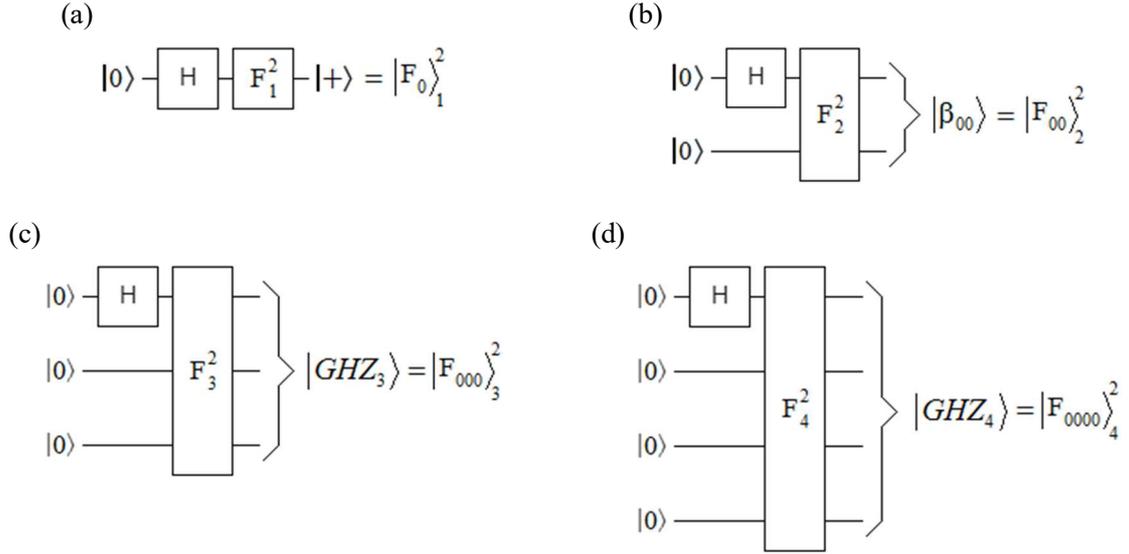

FIG. 18. Quantum stretching: a) for a single qubit, the state $|+\rangle = |F_0\rangle_1^2$ is obtained with $F_1^2 = I_{2^1 \times 2^1}$, b) for two qubits, the output is $|\beta_{00}\rangle = |F_{00}\rangle_2^2$, c) for three qubits, we have $|GHZ_3\rangle = |F_{000}\rangle_3^2$, and for four qubits, the output is $|GHZ_4\rangle = |F_{0000}\rangle_4^2$.

The behavior of Eqs. (77-80) respect to the Hadamard gate (H), can be reproduced with any other gate, e.g., that of Eq.(35), i.e., $\sqrt[4]{X}$ gate, for the state $|\gamma_{00}\rangle = \begin{bmatrix} u & 0 & 0 & v \end{bmatrix}$, with $u = 0.8536+0.3536i$, and $v = 0.1464-0.3536i$, of Table 9. In consequence, the following set of equations arises,

$$|\gamma_0\rangle = F_1^2 \sqrt[4]{X}|0\rangle = u|0\rangle + v|1\rangle$$
$$= \begin{bmatrix} u & v \end{bmatrix}^T \tag{81}$$

$$|\gamma_{00}\rangle = F_2^2 \left(\sqrt[4]{X} \otimes I_{2^1 \times 2^1}\right)|00\rangle = u|00\rangle + v|11\rangle$$
$$= \begin{bmatrix} u & 0 & 0 & v \end{bmatrix}^T \tag{82}$$

$$|\gamma_{000}\rangle = F_3^2 \left(\sqrt[4]{X} \otimes I_{2^2 \times 2^2}\right)|000\rangle = u|000\rangle + v|111\rangle$$
$$= \begin{bmatrix} u & 0 & 0 & 0 & 0 & 0 & 0 & v \end{bmatrix}^T \tag{83}$$

$$|\gamma_{0000}\rangle = F_4^2 \left(\sqrt[4]{X} \otimes I_{2^3 \times 2^3}\right)|0000\rangle = u|0000\rangle + v|1111\rangle$$
$$= \begin{bmatrix} u & 0 & 0 & 0 & 0 & 0 & 0 & 0 & 0 & 0 & 0 & 0 & 0 & 0 & v \end{bmatrix}^T \tag{84}$$

Thus, stretching analysis works for both maximally and non-maximally entangled states.

*Entanglement levels:* Another case of quantum stretching occurs when working also with a gate $F_k^2$ but in a configuration in which its subscript $k$ is not increased, but rather the position of the Hadamard gate (H) at the input of gate $F_k^2$ changes. For example, without losing generality, if we work with the gate of four qubits as $F_4^2$, the sequence of Eqs. (81-84) shows us that, although we work with the same gates, the results are different according to the location of the Hadamard gate (H). This simple process of shifting the Hadamard gate (H) gives rise to a stretching identical to the previous case.

$$|000\rangle|F_0\rangle_1^2 = |000\rangle|+\rangle = F_4^2\left(I_{2^3 \times 2^3} \otimes H\right)|0000\rangle$$
$$= |000\rangle[1 \quad 1]^T/\sqrt{2}$$
(81)

$$|00\rangle|F_{00}\rangle_2^2 = |00\rangle|\beta_{00}\rangle = F_4^2\left(I_{2^2 \times 2^2} \otimes H \otimes I_{2^1 \times 2^1}\right)|0000\rangle$$
$$= |00\rangle[1 \quad 0 \quad 0 \quad 1]^T/\sqrt{2}$$
(82)

$$|0\rangle|F_{000}\rangle_3^2 = |0\rangle|GHZ_3\rangle = F_4^2\left(I_{2^1 \times 2^1} \otimes H \otimes I_{2^2 \times 2^2}\right)|0000\rangle$$
$$= |0\rangle[1 \quad 0 \quad 0 \quad 0 \quad 0 \quad 0 \quad 0 \quad 1]^T/\sqrt{2}$$
(83)

$$|F_{0000}\rangle_4^2 = |GHZ_4\rangle = F_4^2\left(H \otimes I_{2^3 \times 2^3}\right)|0000\rangle$$
$$= [1 \quad 0 \quad 0 \quad 0 \quad 0 \quad 0 \quad 0 \quad 0 \quad 0 \quad 0 \quad 0 \quad 0 \quad 0 \quad 0 \quad 0 \quad 1]^T/\sqrt{2}$$
(84)

In Fig. 19, the transition from $|+\rangle$ to $|GHZ_4\rangle$, passing through $|GHZ_2\rangle$ and $|GHZ_3\rangle$, based on the migration of the Hadamard gate (H) becomes noticeable. This technique can only be carried out with an $F_k^2$-type gate, that is, a QFGS, and not with nested Feynman gates (CNOT). Finally, as in the previous case, if we replace the Hadamard matrix (H) by $\sqrt[4]{X}$, a stretching of the resulting states will be obtained as the gate $\sqrt[4]{X}$ migrates from the lower to the upper qubit. This is how Eqs. (85-88) arise.

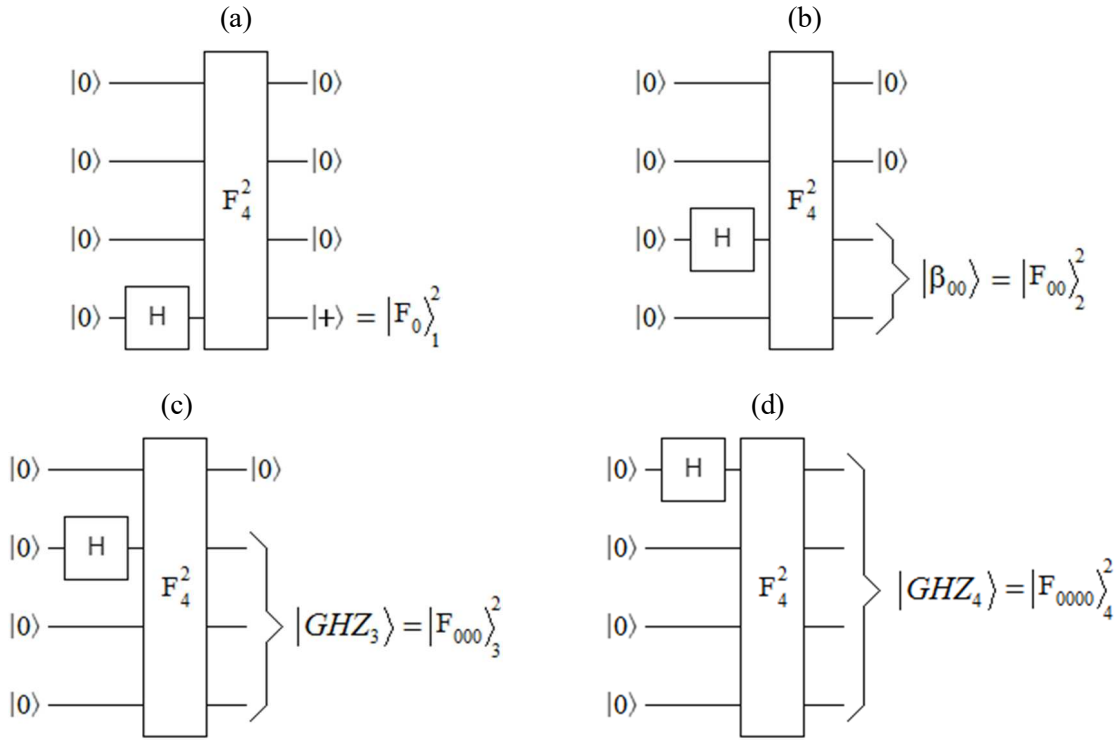

FIG. 19. Entanglement levels: Based on $F_4^2$ and changing the position of the Hadamard matrix (*H*) to its input we go from a) to d), where the outputs are for: a) $|000\rangle|+\rangle$, b) $|00\rangle|\beta_{00}\rangle$, c) $|0\rangle|GHZ_3\rangle$, and d) $|GHZ_4\rangle$.

$$|000\rangle|\gamma_0\rangle = |000\rangle\sqrt[4]{X}|0\rangle = F_4^2\left(I_{2^3\times 2^3}\otimes\sqrt[4]{X}\right)|0000\rangle$$
$$= |000\rangle[u\ \ v]^T \tag{85}$$

$$|00\rangle|\gamma_{00}\rangle = F_4^2\left(I_{2^2\times 2^2}\otimes\sqrt[4]{X}\otimes I_{2^1\times 2^1}\right)|0000\rangle$$
$$= |00\rangle[u\ \ 0\ \ 0\ \ v]^T \tag{86}$$

$$|0\rangle|\gamma_{000}\rangle = F_4^2\left(I_{2^1\times 2^1}\otimes\sqrt[4]{X}\otimes I_{2^2\times 2^2}\right)|0000\rangle$$
$$= |0\rangle[u\ \ 0\ \ 0\ \ 0\ \ 0\ \ 0\ \ 0\ \ v]^T \tag{87}$$

$$|\gamma_{0000}\rangle = F_4^2\left(\sqrt[4]{X}\otimes I_{2^3\times 2^3}\right)|0000\rangle$$
$$= [u\ \ 0\ \ 0\ \ 0\ \ 0\ \ 0\ \ 0\ \ 0\ \ 0\ \ 0\ \ 0\ \ 0\ \ 0\ \ 0\ \ 0\ \ v]^T \tag{88}$$

Thanks to Eqs. (81-84), as well as Eqs. (85-88), it is possible to see that this technique works equally well for both maximally and non-maximally entangled states, respectively.

*Entanglement parallelization:* Based on gates of $F_k^1$-type $\forall\ k$, mutually independent parallel sources of entangled particles can be constructed. For example, without losing generality, for the case of four qubits, we resort to the $F_4^1$ gate, which, together with four *CNOT* gates located in each of its outputs, generates four pairs of entangled particles, i.e., four pairs of the *control-target* type $\{c_i, t_i\}$ ($\forall i \in [0,3]$), where each of these pairs shares a Bell state of type $|\beta_{00}\rangle$. However, they are completely uncorrelated with the members of the other pairs. In Fig. 20, it is possible to identify this configuration. The density matrices between elements of the same pair (control-target), that is, between $c_i$ and $t_i$ ($\forall i \in [0,3]$) is that of Eq.(43), and which we repeat here,

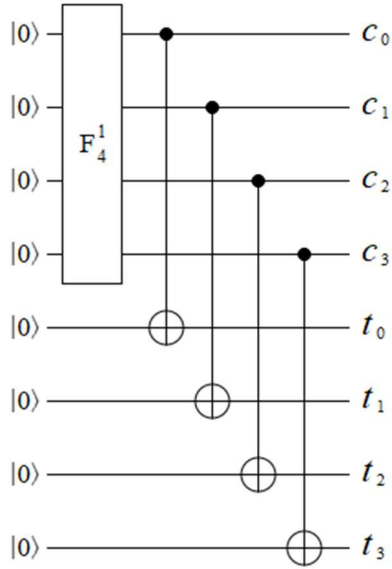

FIG. 20. Entanglement parallelization: Based on a $F_4^1$ gate, each output pair $\{c_i, t_i\}$ ($\forall i \in [0,3]$) shares a state $|\beta_{00}\rangle$ among them, but they are completely uncorrelated with the members of the other pairs.

$$DM_{c_i,t_i} = |\beta_{00}\rangle\langle\beta_{00}| = \frac{1}{2}\begin{bmatrix} 1 & 0 & 0 & 1 \\ 0 & 0 & 0 & 0 \\ 0 & 0 & 0 & 0 \\ 1 & 0 & 0 & 1 \end{bmatrix},\tag{89}$$

while for any other combination of two outputs such that $\forall j \neq i$, the density matrices are:

$$DM_{c_i,t_j} = DM_{c_i,c_j} = DM_{t_i,t_j} = \frac{1}{4}\begin{bmatrix} 1 & 0 & 0 & 0 \\ 0 & 1 & 0 & 0 \\ 0 & 0 & 1 & 0 \\ 0 & 0 & 0 & 1 \end{bmatrix}.\tag{90}$$

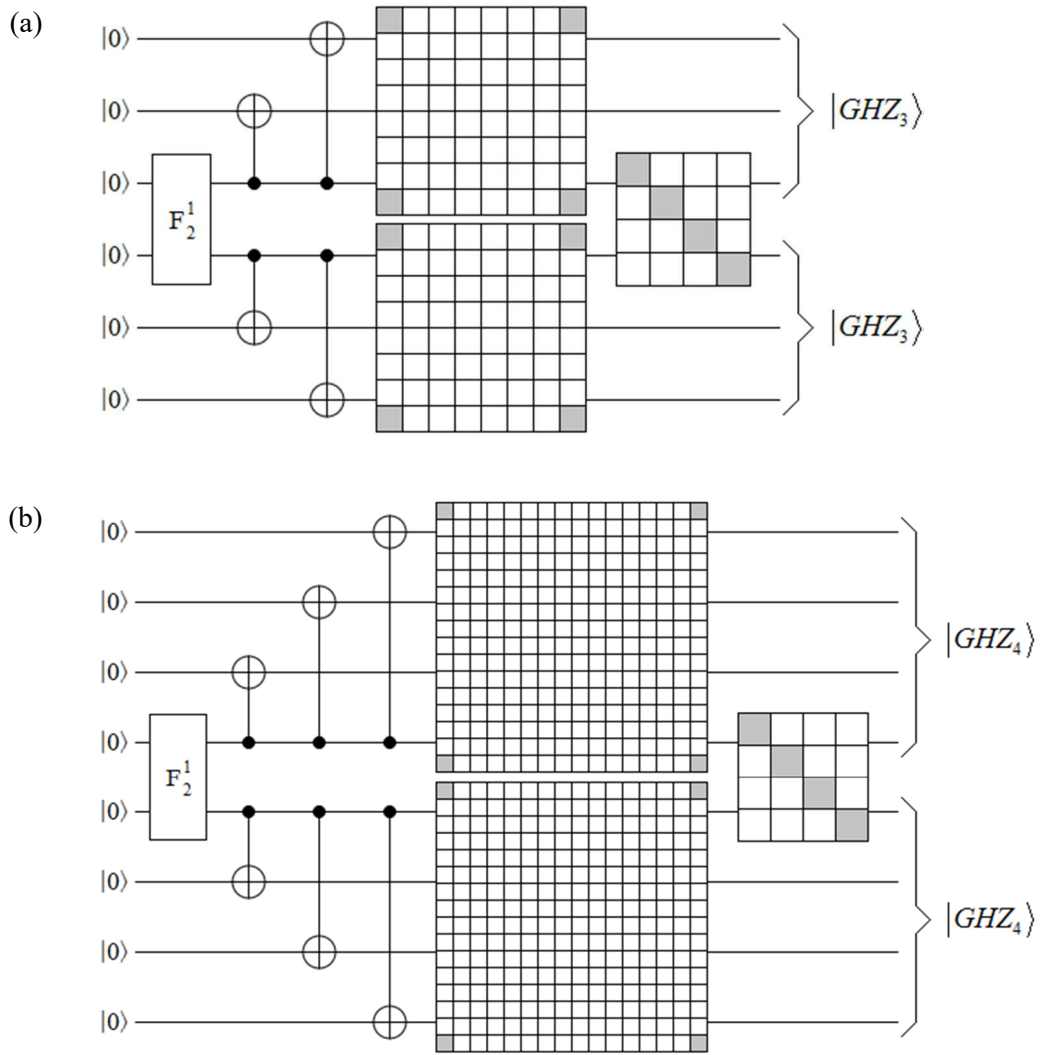

FIG. 21. Configurations of three ($|GHZ_3\rangle$) and four ($|GHZ_4\rangle$) entangled particles based on the $F_2^1$ gate. The grids represent the density matrices, where white elements correspond to a zero, while the gray ones represent values other than zero. Based on these density matrices we can affirm that: a) among qubits of the upper and lower branches there is a state $|GHZ_3\rangle$, but between the two central qubits the correlation is null, b) among qubits of the upper and lower branches there exists a state $|GHZ_4\rangle$, but the correlation is null between the two central qubits, where null correlation means total independence among qubits.

The same situation occurs among groups of three ($|GHZ_3\rangle$) or four ($|GHZ_4\rangle$) particles associated with a gate $F_2^1$, where Fig. 21 represents both configurations. In the case of Fig. 21(a), the top three qubits constitute a $|GHZ_3\rangle$ state, as do the bottom three qubits, however, the middle two qubits are completely decorrelated, i.e., they are independent. Based on Fig. 7(b), we can see the three grids (density matrices) in Fig. 21(a) here too, an element in white means a value equal to zero, while an element in gray represents a value other than zero. Something similar happens in Fig. 21(b), where the top four qubits form a $|GHZ_4\rangle$ state, as do the bottom four qubits, while the middle two qubits have zero correlation. In Fig. 21, the combination of each output of the gate $F_2^1$ with its corresponding CNOT gates gives rise to two parallel sources of entangled pairs of three (a) and four (b) qubits.

Equation (89) shows that the outputs of affine control-target pairs of identical subscript share a Bell state of type $|\beta_{00}\rangle$, while Eq.(90) tells us that, apart from the previous relationship, the outputs are completely independent. This feature makes the configuration of Fig. 20 particularly useful for performing four independent and simultaneous teleportations like those in Fig. 22. This is a natural path from entanglement parallelization to hyper-teleportation. Consequently, this setting controls various spurious effects such as cross-channeling.

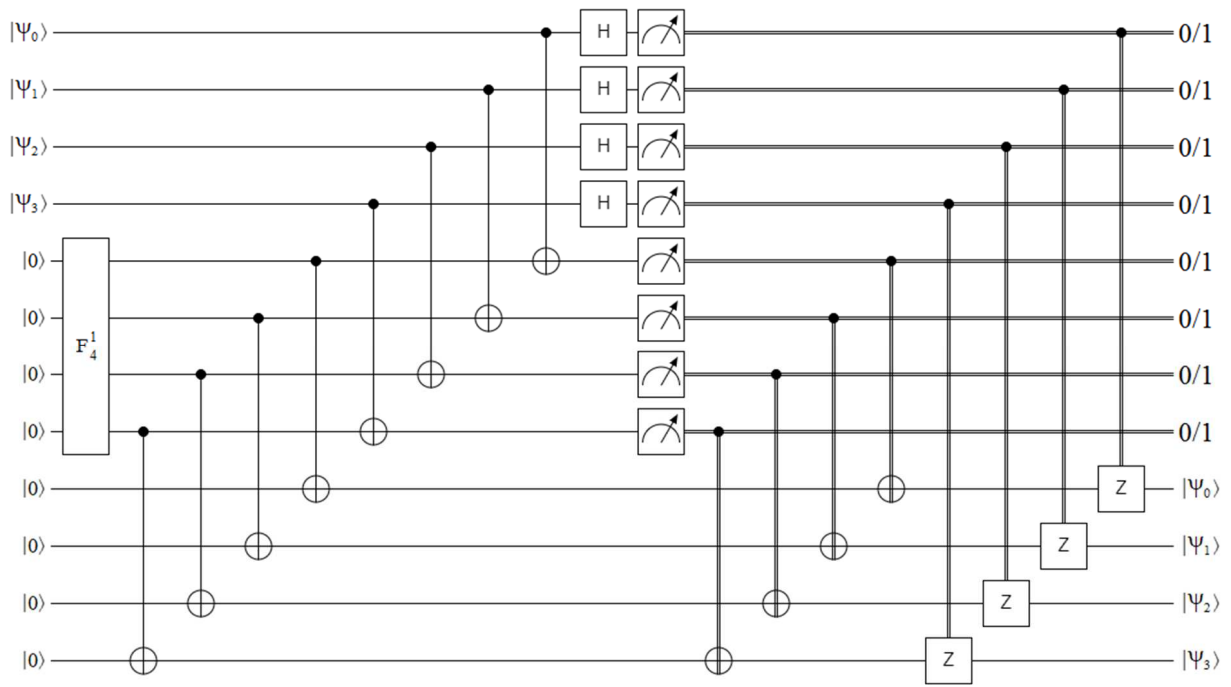

FIG. 22. Based on a configuration as that of Fig. 20, a multiple and simultaneous teleportation of four different qubits at the same time is carried out.

*Quantum secret sharing*[29] (QSS)*:* This protocol constitutes a true central tool inside the quantum cryptography[27] toolbox, and can be interpreted as teleportation for the case of working with sources of 3, 4, and more entangled photons at the same time. The basic scheme of the QSS protocol[3] can be seen in Fig. 23 for the case of working with entangled states of the $|GHZ_3\rangle$ type.

Figure 24 represents a parallel QSS scheme based on two independent sources of $|GHZ_3\rangle$ states, for the simultaneous transmission of two different states, $|\psi_A\rangle$ and $|\psi_B\rangle$, thanks to a configuration that uses only one $F_2^1$ gate. Thanks to $F_2^1$ we can access the first bidirectional QSS protocol in the literature, with a particular projection on the transmission of secure information in the future quantum Internet[9-14].

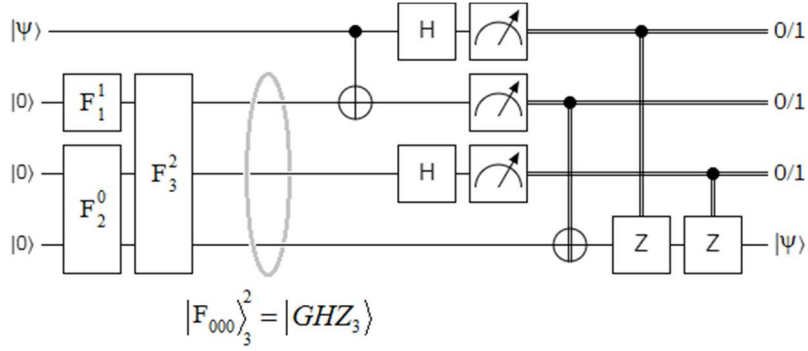

FIG. 23. Quantum secret sharing is based on an entangled source of three particles ($|GHZ_3\rangle$) which is implemented based on quantum Fourier gates like those of Fig. 7(a).

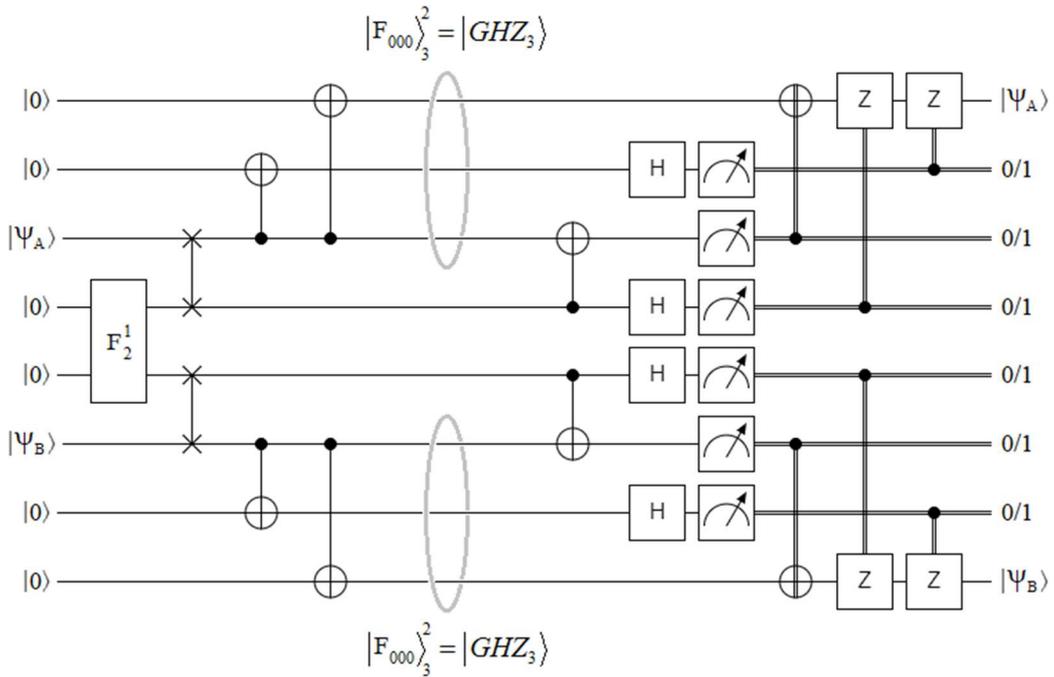

FIG. 24. Parallel quantum secret sharing is based on two independent sources of $|GHZ_3\rangle$ states, for the simultaneous transmission of two different states, $|\psi_A\rangle$ and $|\psi_B\rangle$, thanks to a configuration that uses only one $F_2^1$ gate.

*Quantum repeaters:* Fiber optic cabling for terrestrial implementations of QKD requires quantum repeaters every certain number of kilometers[28], which in turn requires a large amount of quantum memory. The problem is that the key is exposed in its passage through them. There are currently two well-defined lines of research, the first has to do with the development of quantum repeaters that do not require quantum memory, at least not that much, and the second is to replace the same quantum repeaters with some type of implementation based on quantum teleportation[8]. It is imperative to solve this problem to implement the future quantum Internet[9-14].

At this point, two types of quantum repeaters can be identified, those which use:
- Entanglement swapping[30] (transitivity), or
- cascading teleportations (forward).

In previous works, the virtues of quantum Fourier gates (QFG) in the implementation of quantum repeaters based on entanglement swapping were shown[3-6]. Therefore, here quantum repeater based on entanglement swapping with the intervention of entanglement parallelization will be implemented.

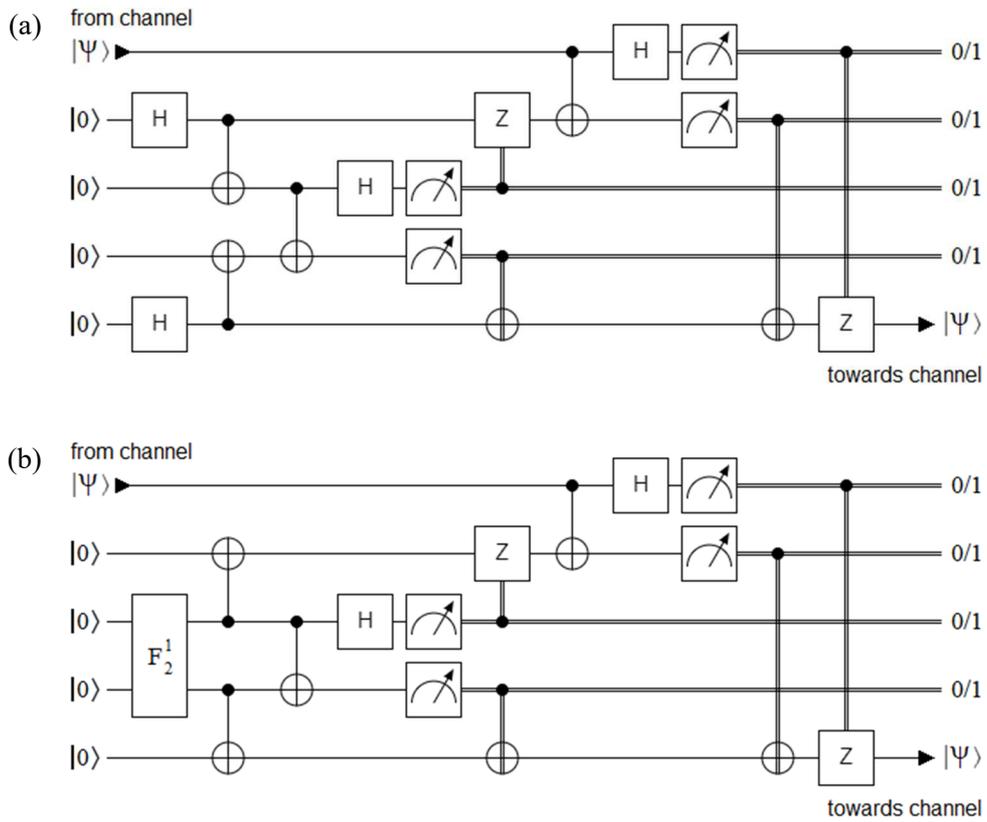

FIG. 25. Quantum repeaters based on entanglement swapping: a) the original version[30], and b) based on a $F_2^1$ gate that generates two parallel and independent Bell states $|\beta_{00}\rangle$.

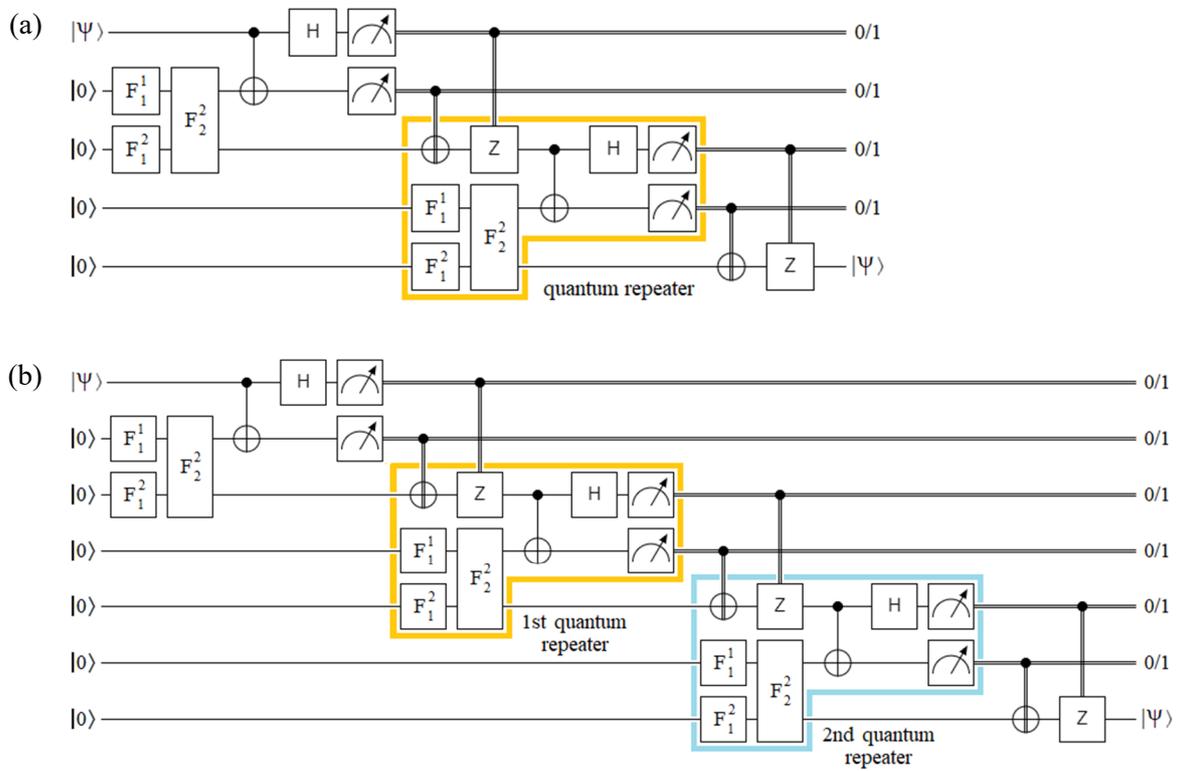

FIG. 26. Quantum repeaters based on teleportation cascade: a) only one quantum repeater and b) two quantum repeaters in consecutive teleportations configuration.

In Fig. 25(a), a classical implementation of a quantum repeater based on entanglement swapping is shown. Links of these structures are regularly used in such a way that a final entanglement is obtained between the first and last link of the chain, which considerably increases the range of the entanglement. For this reason, we speak of transitivity, that is, within the chain of quantum repeaters if A and B are entangled, and B and C are also entangled, eventually A and C will be. Instead, Fig. 25(b) shows the same type of quantum repeater, although the pair of entangled pairs are replaced with a configuration of two sources of entangled particles (independent and parallel) based on the gate $F_2^1$, which puts in evidence another application of entanglement parallelization.

As we have mentioned before, this type of repeater exposes the key that crosses them in a QKD[27] context. For this reason, it is advisable to try other forms of quantum repeaters based on teleportations[8]. In this case, a cascade of teleportations will increase the range of the broadcast. Because of this, the configurations of Fig. 26, which correspond to quantum repeaters based on cascaded quantum teleportations, are proposed. The first uses a single quantum repeater highlighted in orange, see Fig. 26(a), while the second uses two quantum repeaters in cascade, that is, the state received by one is teleported to the next, as can be seen in Fig. 26(b). In the latter case, the second quantum repeater is highlighted in light blue. In both cases of Fig. 26, blocks like those of Fig. 6 and Table 5 are used to generate the states $|\beta_{00}\rangle$.

*Conclusions.-* Another form of entanglement different from those already known and which produces maximally and non-maximally entangled states was presented in this study. The expression rough has to do with the lack of a second layer or block of QFT.

Both the Bell states and the $|GHZ_n\rangle$ states ($\forall n$) are particular cases of the quantum Fourier states. Figures 6-8 demonstrate this. In general, all forms of entanglement are derived from Fourier. In Fig. 2, the Feynman or CNOT gate[1] is expressed as a particular case of quantum Fourier gates.

Quantum teleportation, as well as its projection on the future quantum Internet, exclusively rests on Fourier (the first qubitizer), since all forms of entanglement are just tools in the Fourier toolbox. That is, entanglement is part of something bigger, the best version of it, but that is all, one more member.

The remarkable performance demonstrated by the applications of this technology in the last two sections of this work evidences its projection on QKD[27] and the future quantum Internet[9-14].

**Acknowledgements**.-M.M. thanks the staff of the Knight Foundation School of Computing and Information Sciences at Florida International University for all their help and support.

**Author contributions.**-M.M. conceived the idea and fully developed the theory, wrote the complete manuscript, prepared figures, and reviewed the manuscript.

**Competing Interests**.-M.M. declares that he has no competing interests.

**Additional information.**-Correspondence and requests for materials should be addressed to M.M.

**Data Availability Statement.**-The experimental data that support the findings of this study are available in ResearchGate with the identifier https://doi.org/10.13140/RG.2.2.24802.20161.